\documentclass[12pt]{article}
\usepackage{amsmath}
\usepackage{graphicx,psfrag,epsf}
\usepackage{enumerate}
\usepackage{enumitem}
\usepackage{natbib}
\usepackage{url} % not crucial - just used below for the URL 

\DeclareMathOperator*{\argmax}{arg\,max}
\DeclareMathOperator*{\argmin}{arg\,min}
\usepackage{amsfonts}
\usepackage{verbatim}
\usepackage{xcolor}
\usepackage{amssymb}
\usepackage{epstopdf}%
\usepackage{caption,subcaption}
\usepackage{algorithm,algorithmicx}
\usepackage[noend]{algpseudocode}
\usepackage[detect-all]{siunitx}
\usepackage{tabu}
\usepackage{arydshln}
\usepackage{eufrak}
\usepackage{mathtools}

\newcolumntype{"}{@{\hskip\tabcolsep\vrule width 1pt\hskip\tabcolsep}}
\makeatother
\providecommand{\U}[1]{\protect\rule{.1in}{.1in}}

\newtheorem{theorem}{Theorem}

\newtheorem{definition}{Definition}

\newtheorem{lemma}{Lemma}

\newtheorem{prop}{Proposition}

\newcommand{\bbet}{\boldsymbol{\beta}}
\newcommand{\tbbet}{\boldsymbol{\tilde{\beta}}}
\newcommand{\bbbet}{\boldsymbol{\bar{\beta}}}
\newcommand{\hbbet}{\boldsymbol{\hat{\beta}}}

\newcommand{\bet}{\boldsymbol{\eta}}
\newcommand{\tbet}{\boldsymbol{\tilde{\eta}}}
\newcommand{\hbet}{\boldsymbol{\hat{\eta}}}

\newcommand{\bx}{\mathbf{x}}
\newcommand{\bX}{\mathbf{X}}
\newcommand{\by}{\mathbf{y}}
\newcommand{\bZ}{\mathbf{Z}}
\newcommand{\bz}{\mathbf{z}}
\newcommand{\bu}{\boldsymbol{u}}
\newcommand{\bv}{\boldsymbol{v}}

\newcommand{\calsB}{\mathcal{B}}

\algrenewcommand\algorithmicrequire{\hspace*{\algorithmicindent}\textbf{Input:}}
\algrenewcommand{\algorithmicensure}{\hspace*{\algorithmicindent}\textbf{Initialize:}}
\algdef{SE}[DOWHILE]{Do}{doWhile}{\algorithmicdo}[1]{\algorithmicwhile\ #1}%

\usepackage{placeins}
\usepackage{booktabs}
\usepackage{multirow}
\usepackage{rotating}
\usepackage{afterpage}
\usepackage{setspace}
\usepackage{mathtools}
\usepackage{hyperref}
\hypersetup{
	colorlinks=true,
	linkcolor=blue,
	citecolor=blue,
	filecolor=magenta,
	linktocpage=true,      
	urlcolor=blue,
}
\newcommand{\blind}{0}

\begin{document} 
	
	\onehalfspacing

	\def\spacingset#1{\renewcommand{\baselinestretch}%
		{#1}\small\normalsize} \spacingset{1}

	%%%%%%%%%%%%%%%%%%%%%%%%%%%%%%%%%%%%%%%%%%%%%%%%%%%%%%%%%%%%%%%%%%%%%%%%%%%%%%
	
	\if0\blind
	{
		\title{\bf Robust Multi-Model Subset Selection}
		\author{Anthony-Alexander Christidis\hspace{.2cm}\\
			Department of Statistics, University of British Columbia\\
			Department of Biomedical Informatics, Harvard Medical School\\
			\\
			Gabriela Cohen Freue \\
			Department of Statistics, University of British Columbia}
		\maketitle
	} \fi
	
	\if1\blind
	{
		\bigskip
		\bigskip
		\bigskip
		\begin{center}
			{\LARGE\bf Robust Multi-Model Subset Selection}
		\end{center}
		\medskip
	} \fi

	\begin{abstract}
		Outlying observations can be challenging to handle and adversely affect subsequent analyses, especially in data with increasing dimensional complexity. Although outliers are not always undesired anomalies in the data and may possess valuable insights, only methods that are robust to outliers are able to accurately identify them and resist their influence. In this paper, we propose Robust Multi-Model Subset Selection (RMSS), a method that generates an ensemble of sparse and diverse predictive models that are resistant to outliers. We show that the ensembles generally outperform single-model sparse and robust methods. Cross-validation is used to tune parameters to control levels of sparsity, diversity and robustness. We establish the finite-sample breakdown point of the models generated by RMSS, including that of the Robust Best Subset Selection (RBSS) estimator as a special case. In addition, we develop a tailored computing algorithm to learn the ensemble by leveraging recent developments in  $ \ell_0 $ optimization. Our extensive numerical experiments on synthetic and artificially contaminated real datasets from bioinformatics and cheminformatics demonstrate the competitive advantage of our method over state-of-the-art single-model methods.  The appendix contains all theoretical proofs, additional algorithmic and computational details, and the code and data to reproduce our numerical results.
	\end{abstract}
	
	\noindent%
	\vfill
	
	\newpage
	\spacingset{1.45} % DON'T change the spacing!
	\section{Introduction} \label{sec:introduction}
	
	The rapid growth of digital technologies has led to an explosive increase in data, revolutionizing approaches to modeling and prediction. For example, advances in genomics allow for the simultaneous quantitation of thousands of genes from a patient's sample, which can be used to predict pathogenic processes or responses to therapies \citep[e.g.,][]{byron2016translating}. Large volumes of data of different types, formats, and structures can be rapidly collected, generated and integrated. Thus, modern datasets are often characterized by the presence of a large number of variables (in columns), of which some are irrelevant or redundant, and in general exceeds the number of observations (in rows). Along with more sophisticated processes of obtaining data comes the inclusion of outlying observations, also known as data contamination. In this article, an outlier refers to a case (or row) in the dataset with observed values differing from those of the bulk of the data. These outliers may arise due to errors in data collection, sensitive detection of rare cases, discrepancies in data sources, or data corruption, among other reasons. Although outliers may contain valuable information, only robust methods resistant to outliers can appropriately identify and handle their influence. In this work, we leverage modern computational tools in $ \ell_0 $-optimization to generate an ensemble of sparse and diverse predictive models that are resistant to outlying rows in the dataset and improve the performance of single-model sparse and robust methods.	
	
	Regularized regression methods have been developed to model datasets with many predictors relative to the number of samples, enabling the selection of an optimal subset of predictors for building interpretable predictive models. Prominent examples include the least absolute shrinkage and selection operator (LASSO) \citep{tibshirani1996regression} and the smoothly clipped absolute deviation (SCAD) method \citep{SCAD}. However, depending on the loss function, these methods can be very sensitive to outliers, which adversely affects their variable selection and prediction performance. To address this, robust statistical procedures are derived using loss functions that downweight the effect of outlying observations \citep{maronna2019robust}. In recent years, many proposals have combined regularization and robustness to obtain predictive models resistant to various types of outliers \citep{maronna2011robust, alfons2013sparse, smucler2017robust, cohen2019robust}. Many of these methods have a breakdown point of 50\%, meaning that they can provide a robust prediction when the majority of the cases is not contaminated with outliers.	
	
	Ensemble methods, which generate and aggregate multiple diverse models, often outperform single-model methods in high-dimensional prediction tasks. Traditionally, ensemble methods rely on randomization or some form of heuristics to generate diverse models, and are thus considered ``blackbox" methods. Some notable examples of these ensemble methods include random forests (RF) \citep{RF}, random generalized linear models (RGLM) \citep{random_glm_paper}, gradient boosting \citep{GBM} and all its variations \citep[e.g., ][]{buhlmann2003boosting, chen2016xgboost}. In general, these types of ensemble methods generate a large number of uninterpretable and inaccurate models that are only useful when they are pooled together. More recently,  \cite{christidis2020split} and \cite{christidis2023multimodel}  proposed methods that generate ensembles comprised of a small number of sparse and diverse models learned directly from the data without any form of randomization or heuristics. Each of the models in these ensembles have a high prediction accuracy, and the ensembling of these  small models have been shown to outperform state-of-the-art blackbox ensemble methods on synthetic as well as complex biological and chemical data. However, both these ensembles methods are very sensitive to outliers.

	In this article, we introduce a Robust Multi-Model Subset Selection (RMSS) method to generate ensembles comprised of a small number of sparse, robust, and diverse models in a regression setting. The levels of sparsity, diversity, and robustness are driven directly by the data via cross-validation (CV). We establish the finite-sample breakdown point (FSBP) of the models generated by RMSS, as well as the FSBP of RBSS estimator as a special case. To fit these models, we harness recent developments in $\ell_0$-optimization to develop a tailored computational algorithm with attractive convergence properties. RMSS is shown to outperform state-of-the-art sparse and robust methods in an extensive simulation study and on biological and chemical datasets with simulated contamination. To the best of our knowledge, this is the first robust ensemble method proposed in the literature. Its flexibility may be particularly valuable for practitioners working with moderately high-dimensional data.
	
	The remainder of this article is organized as follows. In Section \ref{sec:lit_review}, we provide a literature review. In Section \ref{sec:RMSS}, we introduce RMSS and some of its special cases. In Section \ref{sec:bp}, we study robustness properties of RBSS and RMSS. In Section \ref{sec:computing}, we provide a computational algorithm to generate RMSS and establish some of its convergence properties. In Section \ref{sec:simulation}, we present a large simulation study. In Section \ref{sec:artificial_contamination}, we apply RMSS on artificially contaminated datasets from bioinformatics and cheminformatics. Concluding remarks are given in Section \ref{sec:summary_future}.

	\section{Literature Review} \label{sec:lit_review}
	In this section, we review a variety of predictive methods proposed in the literature that are related to RMSS and introduce important notation.  
	
	We consider the usual linear regression setting where a dataset comprised of $ n $ observations and $ p $ predictor variables can be used to build a predictive model for a response variable of interest. Let $\mathbf{y}=(y_{1},\dots ,y_{n})^T \in \mathbb{R}^n $, $ \bX \in \mathbb{R}^{n \times p} $, and $\bx_i=(x_{i1},\dots,x_{ip})^T$ be the rows of $\bX$ for $1\leq i \leq n$. We assume a standard linear model
	\begin{equation} \label{eq:true_model}
		y_{i} = \mu + \mathbf{x}_{i}^{T} \boldsymbol{\beta}_{0} + \sigma \epsilon_{i},
		\quad 1\leq i \leq n,
	\end{equation}
	where  $\mu \in \mathbb{R}$ and  $\bbet_{0} \in \mathbb{R}^p$ are the regression coefficients, and the elements of the noise vector $\boldsymbol{\epsilon} = (\epsilon_1, \dots, \epsilon_n)^T \in \mathbb{R}^n$ are independent and identically distributed with mean zero and variance one. We focus our attention on the high-dimensional setting ($ p \gg n $) where the underlying model is sparse, i.e., the number of nonzero elements of the true coefficient vector $ \left\lVert \bbet_0 \right\rVert_0 \ll p $.
	
	\subsection{Single-Model Methods}
	
	Several regression methods were proposed to generate sparse predictive models based on only a subset of the predictor variables, particularly needed when $ p $ is very large compared to the number of observations $n$. The Best Subset Selection (BSS) estimator proposed by \cite{garside1965best} was one of the first variable selection method, which can be defined as the solution to the non-convex (and non-differentiable) minimization problem given by
	\begin{align} \label{eq:BSS}
		\min_{\bbet \in \mathbb{R}^p} \left\lVert \by -\bX \bbet \right\rVert_2^2 \quad \text{subject to} \quad  \left\lVert \bbet \right\rVert_0 \leq t,
	\end{align}
	where $ 0 \leq t \leq \min(n - 1, p) $ is the number of nonzero coefficients, which may be chosen by a model selection criterion \citep[see e.g.,][]{mallows1973some, akaike1974new} or by cross-validation.
	
	Since the BSS optimization problem \eqref{eq:BSS} is an NP-hard problem \citep{welch1982algorithmic}, many sparse regularization methods in the form of convex relaxations of \eqref{eq:BSS} were proposed, such as LASSO, Elastic Net  (EN) \citep{zou2005regularization}, and SCAD \citep{SCAD} methods.
	Although  convex relaxations have much lower computational cost, BSS enjoys better estimation and variable selection properties compared to sparse regularization methods 
	\citep{shen2013constrained}, and often outperforms regularization methods in high-dimensional prediction tasks \citep{hastie2020best}. In an effort to make BSS computational feasible in high-dimensional settings, \cite{bertsimas2016best} proposed  fast and scalable algorithms  to generate solutions to the BSS problem \eqref{eq:BSS} directly.
	
	Since most of these methods are based on the squared loss function, they are  very sensitive to atypical observations in the data, which may adversely affect their variable selection and prediction performances. Over the last two decades, several robust methods have been proposed that can be used when $p \gg n$, can select only a subset of relevant predictors, and are resistant to outliers. \cite{khan2007building} and \cite{khan2007robust} were among the first ones to develop robust stepwise and least angle regression (LARS) \citep{efron2004least} algorithms using robust estimators of pairwise correlations instead of their classical sample counterparts. \cite{alfons2013sparse} introduced a penalized version of the least trimmed squares estimator (LTS) \citep[]{rousseeuw1984least}, called sparseLTS, by minimizing the LASSO-penalized sum of $h$ smallest squared residuals, with $ h \leq \lfloor n/2 \rfloor $. 
	\cite{thompson2022robust} introduced the Robust Best Subset Selection (RBSS) method by combining the LTS loss with the $ \ell_0 $-penalty for the vector of coefficients and developed an efficient algorithm for its computation. The objective function \eqref{eq:BSS} becomes
	\begin{align} \label{eq:RBSS}
		\min_{\mathcal{\substack{\bbet \in \mathbb{R}^p \\ \mathcal{I}' \subseteq \mathcal{I}}}} \; \sum_{i \in \mathcal{I}'} \left(y_i - \bx_i^T \bbet\right)^2 \quad \text{subject to} \quad \begin{cases}
			\left\lVert\bbet\right\rVert_0 \leq t, \\
			|{\mathcal{I}'}| \geq h,
		\end{cases}
	\end{align}
	where $ | \cdot| $ is the cardinality operator, $ \mathcal{I} = \{1, \dots, n\} $ and $h$ is an integer such that $t \leq h \leq n$.
	
	Other sparse robust regression estimators have been later proposed using other loss and penalty functions, including the MM-LASSO \citep{smucler2017robust}, PENSE(M) \citep{cohen2019robust} and their adaptive versions \citep{kepplinger2023robust}.

	\subsection{Ensemble Methods}
	
	Ensemble methods have been proposed to generate and aggregate multiple models with appealing performance in high-dimensional prediction tasks. \cite{ueda1996generalization} decomposed the mean squared prediction error (MSPE) of regression ensembles and showed that the variance of an ensemble is largely determined by how correlated its individual models are. Thus, until recently, most ensemble methods relied on a large number of weak decorrelated models (typically more than $100 $). For example, decorrelation of the individual trees in RF is achieved by random sampling of the data \citep[i.e., bagging,][]{breiman1996bagging} and random sampling of the predictors \citep[i.e., the random predictor subspace method,][]{ho1998random}. 
	Similarly,  ensembles from large number of diverse linear models are generated in the RGLM method \citep{random_glm_paper} and through gradient boosting \citep{chen2016xgboost}. However, their individual models are not interpretable and have weak predictive accuracy. In addition, the selection of predictors is unreliable if randomization is used.
	
	To generate ensembles of sparse, accurate and diverse models, \cite{christidis2020split} and \cite{christidis2023multimodel} relied on the principle of the multiplicity of good models \citep{mccullagh1989monographs}. \cite{christidis2020split} proposed a method called Split-Regularized Regression (SplitReg) that splits the set of predictors into groups and builds a set of sparse models by minimizing an objective function that encourages sparsity within each group and diversity among them. 
	To alleviate using the multi-convex relaxation of SplitReg, \cite{christidis2023multimodel} introduced a multi-model subset selection (MSS) as a generalization of BSS in \eqref{eq:BSS}. Despite their high prediction accuracy,  these methods are very sensitive to outliers in the data. 
	
	\section{Robust Multi-Model Subset Selection} \label{sec:RMSS}
	
	In this section we introduce our Robust Multi-Model Subset Selection (RMSS) estimator, designed to build an ensemble of $G \geq 2$ robust and sparse predictive models from moderately high-dimensional data possibly containing outlying samples. 
	
	Let $\beta_j^g$ denote the coefficient for predictor $j$ in model $g$, for $1 \leq j \leq p$ and $1 \leq g \leq G$. Let $\bbet^g = (\beta_1^g, \dots, \beta_p^g)^T \in \mathbb{R}^p$ be the coefficient vector for model $g$, and let $\bbet_{j\cdot} = (\beta_j^1, \dots, \beta_j^G)^T \in \mathbb{R}^G$ be the vector of coefficients for predictor $j$ across all $G$ models. For a fixed number of models $G \geq 2$, and tuning constants $t, \, u, \, \text{and } h,$ RMSS solves the constrained optimization problem:
	
	\begin{align} \label{eq:RMSS}
		\min_{\substack{\bbet^1, \dots, \bbet^G \in \mathbb{R}^p \\ \mathcal{I}^1, \dots, \mathcal{I}^G \subseteq \mathcal{I}}} \sum_{g=1}^{G} \sum_{i \in \mathcal{I}^{g}} \left(y_i - \bx_i^T \bbet^g\right)^2 \quad \text{subject to} \quad \begin{cases}
			\left\lVert\bbet^g\right\rVert_0 \leq t, \, &1 \leq g \leq G, \\
			\left\lVert\bbet_{j\cdot}\right\rVert_0 \leq u, \, & 1 \leq j \leq p,\\
			|\mathcal{I}^{g}| \geq h, \, & 1 \leq g \leq G.
		\end{cases}
	\end{align}
	Each subset $\mathcal{I}^{g} \subseteq \mathcal{I} = \{1, \dots, n\}$ is comprised of at least $h$ indices of the observations used to estimate model $g$, with $t < h \leq n$. The parameter $h$ controls the robustness of each model by using only the $h$ observations with the smallest sum of squared residuals. The tuning constant $t$ controls the number of predictors selected for each model with $t \leq \min(n-1, p)$. The tuning constant $u$ sets the maximum number of times a predictor can be shared among models with $u \leq G$, thus controlling diversity. In our framework, these constants are chosen by CV (see Section \ref{sec:tuning}).
	
	A nice feature of RMSS is that the individual models in the ensemble may be fit on different subsamples (i.e., different  $\mathcal{I}^g$ subsets). By fitting models on different observation subsets, the ensemble can better exploit uncontaminated portions of the data. Moreover, the combination of this flexible subsampling and variable selection framework, and the multi-model structure of RMSS may help to provide some resistance to cellwise contamination, where individual cells in the data matrix are contaminated \citep{alqallaf2009propagation,raymaekers2024challenges}.  
	
	Since RMSS builds multiple models on different subsets of predictors, contaminated cells in certain observations do not affect all models. On average, each submodel faces a lower effective contamination rate, so that for low levels of cellwise contamination and small model sizes the propagation may be controlled at levels that casewise subsampling-based estimators are able to withstand. The subsampling nature of RMSS further adds flexibility in adapting to different contamination patterns across predictor subsets. Overall, the ensemble can still recover the underlying relationships and the ensemble potentially use more observations relative to single-model robust methods, reducing information loss from the training data. 
	
	We now show that RMSS reduces to several well-known estimators as special cases, depending on the choice of tuning constants.
	
	\begin{prop} \label{prop:rmss_special_cases}
		(Relationship of RMSS to Other Estimators) RMSS generalizes several other subset selection methods:
		\begin{enumerate}
			\item[(I)] If $h=n$, RMSS is equivalent to the non-robust Multi-Model Subset Selection (MSS) method.
			\item[(II)] If $u=G$, the diversity constraint is inactive, and the solution for each of the $G$ models is the optimal solution to the RBSS problem in \eqref{eq:RBSS}.
			\item[(III)] If $u=G$ and $h=n$, RMSS is equivalent to BSS in \eqref{eq:BSS}.
			\item[(IV)] If $u=G$ and $t=p < n-1$, RMSS is equivalent to LTS estimator.
		\end{enumerate}
	\end{prop}
	The proof, given in Appendix A, follows directly from the simplification of the RMSS problem in \eqref{eq:RMSS} under specific parameter settings. For instance, setting $u=G$ renders the diversity constraint inactive, making the optimization problem separable across the $G$ models and thus equivalent to solving the RBSS problem independently for each. The other cases follow from similar arguments. Since the tuning parameters $(t, u, h)$ can be chosen by CV, RMSS can easily adapt to data with different characteristics, including cases with few predictors or no outliers.

	In this article, we generate ensembles using simple model averaging, where the coefficients of the ensemble, $\bbbet$, are the average of the estimated coefficients $\hbbet^g$ of the $G$ models. However, other methods can also be implemented, including weighted model averaging methods \citep{breiman1996stacked} or model aggregation methods \citep{biau2016cobra}.
	
	\section{Finite-Sample Breakdown Point} \label{sec:bp}
	
	This section establishes the finite-sample breakdown point (FSBP) of the RBSS and RMSS coefficient estimators. The FSBP is a standard robustness measure defined by \cite{donoho1983notion} that indicates the smallest fraction of contaminated observations needed to render an estimator meaningless. 
	
	\begin{definition} \label{def:breakdown_point}
		(Finite-Sample Breakdown Point) Let $\bZ = (\bX, \by)$ be a sample of size $n$. The finite-sample breakdown point (FSBP) of an estimator $T$ at the sample $\bZ$ is defined as the smallest fraction of contaminated points $m/n$ that make the estimator's norm unbounded. Formally,
		\begin{align*}
			\varepsilon^*\left(T; \bZ \right) = \frac{1}{n} \min \left\{ m \in \{1, \dots, n\} : \sup_{\tilde{\bZ}_m} \left\lVert T(\tilde{\bZ}_m) \right\rVert_2 = \infty\right\},
		\end{align*}
		where the supremum is taken over all corrupted samples $\tilde{\bZ}_m$ obtained by replacing $m$ points of $\bZ$.
	\end{definition}
	
	While \cite{thompson2022robust} studied the FSBP of the value of the objective function, the FSBP of RBSS coefficient estimator has not been derived. Theorem 1 establishes this result for RBSS, which corresponds to RMSS when $G=1$, as well as for the general case $G \geq 1$ defining RMSS. 
	
	\begin{theorem}
		\label{theo:breakdown_parameters}
		(FSBP of RBSS and Individual RMSS Models) Let $\bZ = (\bX, \by)$ be a sample of size $n$ with $p>1$ covariates (for simplicity, assume a model without intercept). Consider a tuning constant $t$ such that $2t \leq n$ and let $q = \min\{2t, p\}$. Assume that any subset of $k$ covariates, with $t \leq k \leq q$, form a submatrix $\bX_K$ with points in general position. If $\left[ \frac{n+k+1}{2} \right] \le h \le n$, then:
		\begin{enumerate}
			\item[(I)] The finite-sample breakdown point of RBSS coefficient estimator $\hbbet$ is:
			$$ \varepsilon^*\left(\hbbet; \bZ \right) = \frac{n - h + 1}{n}.$$
			
			\item[(II)] The finite-sample breakdown point of each coefficient estimator $\hbbet^g$ of RMSS estimator is:
			$$ \varepsilon^*\left(\hbbet^g; \bZ \right) = \frac{n - h + 1}{n}, \quad 1 \leq g \leq G. $$
			
		\end{enumerate}
	\end{theorem}
	
	The proof of both parts are provided in Appendix B. We first establish the result for the RBSS estimator, a special case of RMSS (Proposition \ref{prop:rmss_special_cases}). Since RBSS reduces to LTS when $t = p < n$, our proof for the lower bound of the FSBP of RBSS builds on the classical LTS breakdown point geometric argument, adapting it to the sparse setting. In our context, the number of predictors $p$ may exceed the sample size $n$ and the estimators are derived under $\ell_{0}$-sparsity constraints, e.g., to have at most $t$ nonzero components. More specifically, for the proofs of the lower bounds of the FSBP of RBSS and RMSS, we adapted Rousseeuw's geometrical construction \citep[]{rousseeuw1984least} by projecting onto the coordinate subspace defined by the selected predictors. The proofs of the upper bounds are constructive and analogous to that of \cite{alfons2013sparse} but adapted to our objective functions with different constraints. 
	
	It is important to note two key differences between the FSBP of LTS and those of RBSS and RMSS. First, the general position condition is more conservative for the sparse case. Since the variables selected by RBSS and RMSS may be affected by the contamination, we require any set of $k$ covariates, with $ t \leq k \leq \min\{2t,p\}$, to have their observations in general position. If $2t \geq p$, and under the condition that $2t < n$, the condition reduces to that of LTS. Second, the FSBP of RBSS and RMSS depend on the ``actual number of parameters" selected by these estimators, rather than on $p$. Interestingly, the dependency of the breakdown point on the effective dimension was also noted by \cite{maronna2011robust} in his analysis of Ridge S- and MM-estimators. 
	
	From Theorem \ref{theo:breakdown_parameters}, if $ h = n $ RBSS and RMSS corresponds to the BSS and MSS, respectively, demonstrating that their breakdown points are $ 1/n $ and thus not resistant to any contamination level. If the trimming parameter $ h $ is chosen by CV, the extent to which RMSS model parameters are resistant to outliers is data-driven.
	
	\section{Computing Algorithm} \label{sec:computing}
	
	A brute-force evaluation of every possible predictor and observation combination in RMSS is computationally infeasible, as shown by a combinatorial analysis in Appendix C. We therefore propose a practical algorithm that searches over a three-dimensional grid of the tuning parameters $(t, u, h)$ in \eqref{eq:RMSS}.
	
	Our procedure begins by robustly standardizing the data using medians and median absolute deviations. We then generate initial disjoint predictor subsets (Section \ref{sec:initial_subsets}) to serve as a ``warm start" for an algorithm that computes solutions over the entire $(t, u, h)$ grid (Section \ref{sec:grid_solutions}). While we also developed an optional neighborhood search to refine these solutions (see Appendix F), its marginal impact on performance did not justify its computational cost for the main procedure. Finally, all estimated coefficients are returned to their original scale.
	
	\subsection{Initial Predictor Subsets} \label{sec:initial_subsets}
	
	To initialize the main RMSS algorithm, we require a high-quality, robust partition of predictors into $G$ disjoint subsets. \cite{christidis2023multimodel} proposed a multi-model forward selection algorithm that selects $ G $ disjoint subsets of predictors to initialize the algorithm of MSS. However, their approach is sensitive to outliers in the data. Thus, in this section we propose a robust multi-model stepwise selection procedure to select disjoint initial subsets of predictors. 
	
	Algorithm \ref{alg:stepwise_algo} generalizes the robust forward stepwise regression algorithm of \cite{khan2007building} to multiple models. For a single-model framework, they showed that the stepwise forward search algorithm depends only the sample means, variances and correlations between the variables in the model. Thus, to make the algorithm robust to outlying observations, they proposed replacing these sample estimators by robust counterparts to compute the residual sum of squares and partial $F$-rules. In Appendix D, we explicitly show this connection, which in turn enabled us to derive the computationally efficient and novel recursive formulas shown in Proposition \ref{prop:optimal_var_formulas}. Consistently with our algorithm and code, these formulas specify the rule for selecting the best single predictor from the candidate set.
	
	\begin{prop} \label{prop:optimal_var_formulas}
		(Correlation-Based Formulas)
		Let $\mathbf{\hat{r}_{y}} = (\hat{r}_{y1}, \dots, \hat{r}_{yp})^T \in \mathbb{R}^p$ be the vector of robust correlations between the response and predictors, and let $\mathbf{\hat{\Sigma}} \in \mathbb{R}^{p \times p}$ be the robust correlation matrix of the predictors. For a single model $g$ with current active set $\mathcal{J}^g \subseteq \mathcal{J} = \{1, \dots, p\}$, the following rules determine how the optimal $k$-th predictor, $j_k$, is selected from a candidate set $\mathcal{C} \subseteq \mathcal{J}$ and how the robust Residual Sum of Squares (rRSS) is calculated:
		\begin{itemize}
			\item If the model is empty ($k=1$ and $\mathcal{J}^{g} = \emptyset$):
			\begin{align*}
				j_1 = \argmax_{j \in \mathcal{C}} |\hat{r}_{yj}|, \qquad \text{rRSS}_1 = n(1 - \hat{r}_{yj_1}^2).
			\end{align*}
			\item If the model is not empty ($k \geq 2$ and $\mathcal{J}^{g}=\{j_1,\dots,j_{k-1}\}$):
			\begin{align*}
				j_k = \argmax_{j \in \mathcal{C}} \left| \frac{P_{yj.j_1 \ldots j_{k-1}}}{\sqrt{P_{jj.j_1 \ldots j_{k-1}}}} \right|, \qquad \text{rRSS}_k = \text{rRSS}_{k-1} - \frac{P_{yj_k.j_1 \ldots j_{k-1}}^2}{P_{j_k j_k.j_1 \ldots j_{k-1}}}.
			\end{align*}
		\end{itemize}
		The partial covariance terms $P$ are computed recursively as follows.
		\begin{equation*}
			\begin{minipage}[t]{0.45\textwidth}
				\centering
				{For $k=2$:}
				\begin{align*}
					\begin{cases}
						P_{yj.j_1} = n \left( \hat{r}_{yj} - \hat{\Sigma}_{jj_1} \hat{r}_{yj_1} \right), \\
						P_{j_a j_b.j_1} = n \left( \hat{\Sigma}_{j_a j_b} - \hat{\Sigma}_{j_a j_1} \hat{\Sigma}_{j_b j_1} \right).
					\end{cases}
				\end{align*}
			\end{minipage}
			\hfill
			\begin{minipage}[t]{0.5\textwidth}
				\centering
				{For $k \geq 3$:}
				\begin{align*}
					\begin{cases}
						P_{yj.j_1 \ldots j_k} = P_{yj \ldots j_{k-1}} - \frac{P_{jj_k \ldots j_{k-1}}}{P_{j_k j_k \ldots j_{k-1}}} P_{yj_k \ldots j_{k-1}}, \\
						P_{j_a j_b \ldots j_k} = P_{j_a j_b \ldots j_{k-1}} - \frac{P_{j_a j_k \ldots j_{k-1}} P_{j_b j_k \ldots j_{k-1}}}{P_{j_k j_k \ldots j_{k-1}}}.
					\end{cases}
				\end{align*}
			\end{minipage}
		\end{equation*}
	\end{prop}
	
	Algorithm \ref{alg:stepwise_algo} describes our multi-model initialization procedure, which uses the formulas from Proposition \ref{prop:optimal_var_formulas} within a competitive framework. The goal of this framework is to produce disjoint initial predictor sets, which encourages diversity and provides a more effective ``warm start" for the main RMSS optimization. 
	
	In our implementation, we compute robust correlation estimates via the state-of-the-art Detect Deviating Cells (DDC) method of \cite{rousseeuw2018detecting}, which can be scaled to high-dimensional settings using properties of product moments \citep{raymaekers2021fast}. Following in \cite{khan2007building}, Proposition \ref{prop:optimal_var_formulas} shows how these robust estimates are used to compute robust residual sum of squares (rRSS) and robust partial $F$-rules. Due to post-inference problems and the use of plug-in estimators, the resulting methodology can not be used to perform formal $F$-tests with valid $p$-values from the $F$-distribution. In Algorithm \ref{alg:stepwise_algo} these rules and $p$-values are only used to develop a selection and a stopping criterion. 
	
	At each iteration, the algorithm employs a greedy strategy: it identifies the best candidate for each model, then assigns the single predictor with the smallest robust partial $F$-test $p$-value across all models to its corresponding model, removing it from the candidate pool $\mathcal{C}$. The saturation threshold, $\gamma \in (0,1)$, controls variable addition during initialization. While a selective approach (e.g., $\gamma = 0.05$) is reasonable, an inclusive strategy is also justified. Since the primary objective is to generate a rich warm start and the final model sparsity is rigorously controlled later by the tuning parameter $t$, a non-restrictive value (e.g., $\gamma = 1$) can be used. This allows the stepwise procedure to populate the initial models generously until they reach their maximum size or the predictor pool is exhausted.
	
	\begin{algorithm}[ht!]
		\caption{Robust Multi-Model Stepwise Selection \label{alg:stepwise_algo}}
		\begin{algorithmic}[1]
			\Require{Robust correlation vector of the response $ \mathbf{\hat{r}_{y}} \in \mathbb{R}^p $, robust correlation matrix of predictors $ \mathbf{\hat{\Sigma}} \in \mathbb{R}^{p \times p} $, number of models $ G \geq 2 $, and saturation threshold $ \gamma \in (0,1) $.}
			\vspace{0.2cm}
			\Ensure{The set of candidates $ \mathcal{C} = \{1, \dots, p\} $, and for each model the set of model predictors $ \mathcal{J}^{g} = \emptyset $ and the model saturation indicator $ M_g = \textsc{false} $, $ 1 \leq g \leq G $.}
			\Statex
			\State Repeat the following steps until $ M_g = \textsc{true} $ for all $g$, $ 1 \leq g \leq G $: \label{alg1:step1}
			\begin{enumerate}[label*=\footnotesize 1.\arabic*.]
				\item For each model $ g $ satisfying $ M_g=\textsc{false} $: 
				\begin{enumerate}[label*=\footnotesize \arabic*:]
					\item Using Proposition \ref{prop:optimal_var_formulas} with candidate set $\mathcal{C} $, identify candidate predictor $ j_{g} $ that maximizes the decrease in rRSS.
					\item Calculate the $ p $-value $ \gamma_g $ from the robust partial $ F $-rule using the decrease in rRSS for adding predictor $ j_{g} $.
					\item If $ \gamma_g \geq \gamma $, set $ M_g = \textsc{true} $.
				\end{enumerate}
				\item Identify the unsaturated model $ g^* $ with the smallest $ p $-value $ \gamma_{g^*} $.            
				\item If $ \gamma_{g^*} < \gamma $:
				\begin{enumerate}[label*=\footnotesize \arabic*:]
					\item Update the set of predictors for model $ g^* $: $ \mathcal{J}^{g^*} \leftarrow \mathcal{J}^{g^*} \cup \{j_{g^*}\} $.
					\item Update the set of candidate predictors $ \mathcal{C} \leftarrow \mathcal{J} \setminus \{j_{g^*}\} $.
					\item If $ |\mathcal{J}^{g^*}| = n-1 $, set $ M_{g^*}=\textsc{true} $.
				\end{enumerate}         
			\end{enumerate}
			\State Return the sets of model predictors $ \mathcal{J}^{g} $, $ 1 \leq g \leq G $.
		\end{algorithmic}
	\end{algorithm}

	\subsection{Computing RMSS over a Grid of Tuning Constants} \label{sec:grid_solutions}
	
	To develop an efficient computing algorithm to compute RMSS over a grid of $ t $, $ u $ and $ h $ values, we recast \eqref{eq:RMSS} in its equivalent form using auxiliary variables $ \boldsymbol{\eta}^g = (\eta_1^g,\dots,\eta_n^g )^T \in \mathbb{R}^n$, $ 1 \leq g \leq G $, with $0$ entries for the smallest $h$ residuals of each model $g$, and the corresponding residuals otherwise. For any $ t $, $ u $ and $ h $ values, the equivalent reformulation of RMSS is given by
	
	\begin{align} \label{eq:RBSpS-aux}
		\min_{\substack{\bbet^1, \dots, \, \bbet^G \in \mathbb{R}^p \\ \boldsymbol{\eta}^1, \dots, \, \boldsymbol{\eta}^G \in \mathbb{R}^p}} \;   \; \sum_{g=1}^{G} \mathcal{L}_n\left(\bbet^g, \boldsymbol{\eta}^g \vert \by, \bX\right) \quad \text{subject to} \quad \begin{cases}
			\left\lVert\bbet^g\right\rVert_0 \leq t, \, &1 \leq g \leq G, \\
			\left\lVert\bbet_{j\cdot}\right\rVert_0 \leq u, \, & 1 \leq j \leq p,\\
			\left\lVert\boldsymbol{\eta}^g\right\rVert_0 \leq n - h, \, & 1 \leq g \leq G,
		\end{cases}
	\end{align}
	where the loss function $\mathcal{L}_n$ used for each model $g$ is given by
	\begin{align}
		\mathcal{L}_n\left(\bbet^g , \boldsymbol{\eta}^g  \vert \by, \bX\right) = \left\lVert \by - \bX \bbet^g  - \boldsymbol{\eta}^g  \right\rVert_2^2.
	\end{align}
	% and the gradients of $\mathcal{L}_n$ with respect to $ \bbet $ and $ \boldsymbol{\eta} $ are given by
	% \begin{align}
		% 		\nabla_{\bbet} \mathcal{L}_n\left(\bbet,\boldsymbol{\eta} | \by, \bX \right) &= 2\bX^T \left(\bX \bbet + \boldsymbol{\eta} - \by \right), \\
		% 			\nabla_{\boldsymbol{\eta}} \mathcal{L}_n\left(\bbet, \boldsymbol{\eta} | \by, \bX \right) &= 2  \left(\bX \bbet + \boldsymbol{\eta} - \by\right).
		% \end{align}
	The gradients %of $\mathcal{L}_n$ with respect to any $ \bbet $ and $ \boldsymbol{\eta} $, 
	$ \nabla_{\bbet} \mathcal{L}_n(\bbet,\boldsymbol{\eta} | \by, \bX ) $ and $ \nabla_{\boldsymbol{\eta}} \mathcal{L}_n(\bbet, \boldsymbol{\eta} | \by, \bX ) $ are both Lipschitz continuous with Lipschitz constants $ \ell_{\bbet} = 2  \lVert\bX^T \bX \rVert_2^2 $ and $ \ell_{\boldsymbol{\eta}} = 2$, respectively (see proofs in Appendix E).
	
	%	Essential to our computing algorithm are  two projection operators which we define below.
	We now introduce some notation needed to outline our computing algorithms.
	\begin{definition}
		(Projection Operator) For any $ v \in \mathbb{R}^p $ and $ r \in \mathbb{R} $, the projection operator $ \mathcal{P}(v; r ) $, defined as
		\begin{align}
			\mathcal{P}\left(v; r \right) \in \argmin_{w \in \mathbb{R}^p}  \left\lVert w - v \right\rVert_2^2 \quad \text{subject to} \quad 
			\left\lVert w  \right\rVert_0 \leq r 
		\end{align}
		retains the $ r $ largest elements in absolute value of the vector $ v $.
	\end{definition}
	\begin{definition}
		(Projected Subset Operator) For any $ v \in \mathbb{R}^p $, $\mathcal{S} \subseteq \mathcal{J} = \{1, \dots, p\}$ and $ r \in \mathbb{R} $, the projected subset operator $ \mathcal{Q}(v;  \mathcal{S}, r )$, defined as
		\begin{align}
			\mathcal{Q}\left(v;  \mathcal{S}, r \right) \in \argmin_{w \in \mathbb{R}^p} \; \left\lVert w - v \right\rVert_2^2 \quad \text{subject to} \quad \begin{cases}
				\left\lVert w  \right\rVert_0 \leq r \\ 
				\{j \in \mathcal{J}: w_j \neq 0\} \subseteq \mathcal{S} 
			\end{cases}
		\end{align}
		retains the $ r $ largest elements in absolute value of the vector $ v $ that belong to the subset $ \mathcal{S} $.
	\end{definition}
	%	The operator $ \mathcal{P}(v; r ) $ retains the $ r $ largest elements in absolute value of the vector $ v $, while the operator $ \mathcal{Q}(v;  S, r ) $ retains the $ r $ largest elements in absolute value of the vector $ v $ that belong to the set $ S $. 
	
	Note that both $ \mathcal{P}(v; r ) $ and $ \mathcal{Q}(v; \mathcal{S}, r ) $ are set-valued maps since more than one possible permutation of the indices $ \mathcal{J} = \{1, \dots, p\} $ and $ \{j \in \mathcal{J}: j \in \mathcal{S}\} $ may exist.

	For a given set of multi-model coefficient estimates $ \hbbet^g = (\hat{\beta}_1^g, \dots, \hat{\beta}_p^g)^T \in \mathbb{R}^p $, $ 1 \leq g \leq G $, let $ \mathcal{J}^{g}=\{j \in \mathcal{J}: \hbbet_j^g \neq 0 \} $ be the subsets of predictors' indices included in each model and $\mathcal{S}_u^{(g)}$ be the subsets of predictors' indices used in at most $ u - 1 $ models excluding model $ g $,
	
	\begin{align} \label{eq:set_allowed}
		\mathcal{S}_u^{g} = \left\{j \in \mathcal{J}: \sum_{\substack{l=1 \\ l \neq g}}^G \mathbb{I}\left(j \in \mathcal{J}^{l}\right) \leq u-1\right\}.
	\end{align}
	
	\noindent Lastly, denote $ \bX_{\mathcal{S}} \in \mathbb{R}^{n \times |\mathcal{S}|} $  the submatrix of $ \bX $ with column indices $ \mathcal{S} \subseteq \mathcal{J} = \{1, \dots, p\}$.
	
	For fixed tuning constants $ t $, $ u $ and $ h $ and given some starting values $ (\tbbet^g, \tbet^g)$, $ 1 \leq g \leq G $, Algorithm \ref{alg:projected_algo},  outlines the steps to perform a projected subset block gradient descent (PSBGD) to generate robust multi-model estimates. For each model at a time, the algorithm alternates between updates of the coefficient estimates $ \hbbet^g $ and updates of the subsamples selected given the size of the updated residuals in $ \hbet^g $ until convergence is achieved. 
	
	\begin{algorithm}[ht!]
		\caption{\label{alg:projected_algo} Projected Subset Block Gradient Descent (PSBGD)}
		\begin{algorithmic}[1]
			\Require{Matrix of predictor variables $\bX \in \mathbb{R}^{n \times p}$, response vector $\by \in \mathbb{R}^n$,  starting values  $ (\tbbet^g, \tbet^g)$, $ 1 \leq g \leq G $, tuning parameters $ t $, $ u $ and $ h $, and tolerance parameter $\epsilon>0$.}
			\Ensure{For each model $g$, the set of model predictors $\mathcal{J}^{g} = \{j \in \mathcal{J}: \tbbet_j^g \neq 0 \} $, $ 1 \leq g \leq G $.}
			% \Statex
			%	\State Initialize the sets of model predictors $ \mathcal{J}^{g} = \{j \in J: \tbbet_j^g \neq 0 \} $, $ 1 \leq g \leq G $. \label{alg2:step1}
			\Statex
			\State Repeat the following steps for each model $ g $, $ 1 \leq g \leq G $: \label{alg2:step2}
			\begin{enumerate}[label*=\footnotesize 1.\arabic*:]
				\item[\footnotesize 1.1:] Create the set of predictors' indices $ \mathcal{S}_u^{g} $ via \eqref{eq:set_allowed}
				and compute the Lipschitz constant $ \ell_{\bbet^{g}} = 2 \lVert \bX_{\mathcal{S}_u^{g}}^T \bX_{\mathcal{S}_u^{g}} \rVert_2  $.
				\item[\footnotesize 1.2:] Update, sequentially, current estimates $ \tbbet^g$ and $\boldsymbol{\tilde{\eta}}^g $ via 
				\begin{align*}
					\hbbet^g &\in \mathcal{Q}\left(\tbbet^g - \frac{1}{L_{\bbet^{g}} } {\nabla}_{\bbet} \mathcal{L}_n\left(\bbet, \boldsymbol{\hat{\eta}}^g|\by, \bX\right)\Big\vert_{\bbet = \tbbet^g}; \, \mathcal{S}_u^{g}, r \right) \\
					\hbet^g &\in \mathcal{P}\left(\tbet^g - \frac{1}{L_{\bet} } {\nabla}_{\bet} \mathcal{L}_n\left(\hbbet^g,\bet|\by, \bX\right)\Big\vert_{\bet = \tbet^g}; n - h \right)
				\end{align*}
				with $L_{\bbet^{g}} \geq \ell_{\bbet^{g}} $ and $ L_{\boldsymbol{\eta}} \geq \ell_{\boldsymbol{\eta}} $, and repeat until $\mathcal{L}_n(\tbbet^g, \boldsymbol{\tilde{\eta}}^g|\by, \bX) - \mathcal{L}_n(\hbbet^g,  \boldsymbol{\hat{\eta}}^g|\by, \bX) \leq \epsilon$.
				\item[\footnotesize 1.3:] Update the model predictors $ \mathcal{J}^{g} = \{j \in \mathcal{J}: \hat{\beta}_j^g \neq 0 \} $ and compute the set of model subsamples $ \mathcal{I}^{g} = \{i \in \mathcal{I}: \hat{\eta}_i^g =0\} $.
				\item[\footnotesize 1.4:] Compute the final model coefficients
				\begin{align*}
					\hbbet^g = \argmin_{\substack{\bbet \in \mathbb{R}^p}} \sum_{i \in \mathcal{I}^{g}} \left(y_i - \bx_i^T \bbet\right)^2 \quad \text{subject to} \quad \beta_j = 0, j \notin \mathcal{J}^{g}, \quad 1 \leq g \leq G.
				\end{align*}
			\end{enumerate}
			\State Return the pairs $ (\hbbet^g, \hbet^g) $, $ 1 \leq g \leq G $.
		\end{algorithmic}
	\end{algorithm}
	
	Proposition \ref{convergence} below establishes the convergence of the Algorithm \ref{alg:projected_algo} (see the proof in Appendix E).
	
	\begin{prop}\label{convergence}
		(Model-Wise Convergence of PSBGD Algorithm) For each model $ g $ in \eqref{eq:RBSpS-aux}, step \ref{alg2:step2} of Algorithm  \ref{alg:projected_algo} generates a converging sequence for the pair ($ \hbbet^g $, $ \hbet^g $) and the inequalities
		\begin{align*}
			&\left\lVert \hbbet^g - \mathcal{Q}\left(\hbbet^g - \frac{1}{L_{\bbet^{g}} } {\nabla}_{\bbet} \mathcal{L}_n\left(\bbet, \boldsymbol{\hat{\eta}}^g|\by, \bX\right)\Big\vert_{\bbet = \hbbet^g}; \, \mathcal{S}_u^{g}, r \right)\right\rVert_2^2 \leq \delta,   \\
			&\left\lVert \hbet^g - \mathcal{P}\left(\hbet^g - \frac{1}{L_{\bet} } {\nabla}_{\bet} \mathcal{L}_n\left(\hbbet^g,\bet|\by, \bX\right)\Big\vert_{\bet = \hbet^g}; n - h \right) \right\rVert_2^2 \leq \delta,
		\end{align*}
		can be achieved in $ O(1/\delta) $ iterations.
	\end{prop}
	
	%	Algorithm \ref{alg:projected_algo} generates solutions for fixed tuning parameters $ t $, $ u $ and $ h $ given starting values. 
	
	In Algorithm \ref{alg:decrementing_projected_algo}, we outline the steps to generate the solutions $(\hbbet^g[t, u, h], \hbet^g[t, u, h]) $, $ 1 \leq g \leq G $,  for any $ t $, $ u $, and $ h $ over the grids $ \mathcal{T} = \{t_1, \dots, t_q \} $, $ \mathcal{U} = \{1, \dots, G\} $ and $ \mathcal{H} = \{h_1, \dots, h_r\} $, respectively. 
	
	\begin{algorithm}[ht!]
		\caption{Decrementing Diversity PSBGD \label{alg:decrementing_projected_algo}}
		\begin{algorithmic}[1]
			\Require{Design matrix $\bX \in \mathbb{R}^{n \times p}$, response vector $\by \in \mathbb{R}^n$, grids of tuning parameters $ \mathcal{T} = \{t_1, \dots, t_q \} $, $ \mathcal{U} = \{1, \dots, G\} $ and $ \mathcal{H} = \{h_1, \dots, h_r\} $, and tolerance parameter $\epsilon>0$.}
			\Statex
			\State Use Algorithm \ref{alg:stepwise_algo} to obtain the sets of predictors' indices $ \mathcal{J}^{g} $, $ 1 \leq g \leq G $,  and compute the initial model coefficients 
			\begin{align*}
				\hbbet_{\text{init}}^g = \argmin_{\substack{\bbet \in \mathbb{R}^p}} \left\lVert \by - \bX  \bbet  \right\rVert_2^2 \quad \text{subject to} \quad \beta_j = 0, j \notin \mathcal{J}^{g}.
			\end{align*}
			\State For each combination of  $ t \in \mathcal{T} $ and $ h \in \mathcal{H} $:
			\begin{enumerate}
				\item[\footnotesize 2.1:] Compute the pairs $ (\hbbet^g[t, 1, h], \hbet^g[t, 1, h]) $, $ 1 \leq g \leq G $, using Algorithm \ref{alg:projected_algo} initialized with $ (\hbbet_{\text{init}}^g, \mathbf{0}_n)$ where $ \mathbf{0}_n = (0, \dots, 0)^T \in \mathbb{R}^n $,  $ 1 \leq g \leq G $.
				\item[\footnotesize 2.2:] For $ u=2,\dots, G $:
				\begin{itemize}
					\item[\footnotesize 2.2.1:] Compute the pairs $ (\hbbet^g[t, u, h], \hbet^g[t, u, h]) $, $ 1 \leq g \leq G $,  using Algorithm \ref{alg:projected_algo} initialized with $ (\hbbet^g[t, u -1, h], \hbet^g[t, u -1, h]) $, $ 1 \leq g \leq G $.
				\end{itemize}
			\end{enumerate}
			\State Return the pairs  $ (\hbbet^g[t, u, h], \hbet^g[t, u, h]) $, $ 1 \leq g \leq G $, for all combinations of $ t \in \mathcal{T} $, $ u \in \mathcal{U}$ and $ h \in \mathcal{H} $. 
		\end{algorithmic}
	\end{algorithm}	
	
	\subsection{Selection of Tuning Parameters}\label{sec:tuning}
	
	We use 5-fold CV to select the final combination of $(t, u, h)$ from the candidate grids $\mathcal{T}, \mathcal{U},$ and $\mathcal{H}$. Since the test folds may contain outliers, we use the fast and robust $\tau$-scale estimator of prediction residuals to evaluate performance for each parameter combination \citep{maronna2002robust}. Our final model is the one corresponding to the combination with the smallest robust scale estimate. This robust CV approach provides stable parameter selection even in high-noise scenarios where traditional CV using mean squared error may fail. As previously mentioned, the combination of multiple models ($G > 1$) with the diversity constraint ($u < G$) provides natural regularization that helps prevent overfitting and enhances stability, which is reflected in the strong performance of RMSS across a range of signal-to-noise scenarios in our simulations.

	The tuning parameters in RMSS have clear interpretations that guide their practical selection. The subsample size $h$, controls the robustness of each model by using only the $h$ observations that produce the smallest sum of squared residuals. This tuning parameter can be set based on the anticipated level of contamination in the data (for example, $h = 0.75n$ corresponds to 25\% contamination) or chosen by CV from a grid (e.g., $\mathcal{H} = \{\lfloor n/2 \rfloor + 1, \dots, n\}$). The sparsity parameter, $t$, controls model complexity for each model with $t \leq \min(n-1, p)$ and less dense grids can be used to speed computation. The diversity parameter, $u$, controls the number of predictors shared across models, where $u = 1$ enforces maximum diversity. Typically, $u = 1$ (completely distinct models) or $u = 2$ (predictors shared at most by $2$ models) perform well in our simulation studies. 
	
	For the ensemble size $G$, we recommend starting with $G = 5$ for $p \ll 500$ and $G = 10$ otherwise, as our simulations show diminishing returns beyond these choices due to increased computational cost without proportional performance gains. The CV procedure scales as $O(G \times \text{CV-folds} \times |\mathcal{T}| \times |\mathcal{U}| \times |\mathcal{H}|)$, making it computationally feasible for typical problems with modest grid sizes.
	
	\subsection{Software}
	
	The implementation of robust multi-model stepwise selection outlined in Algorithm \ref{alg:stepwise_algo} is available on CRAN \citep{CRAN} in the \texttt{R} package \texttt{robStepSplitReg} \citep{robStepSplitReg_package}. 
	The implementation of Algorithms \ref{alg:projected_algo} - \ref{alg:decrementing_projected_algo} to fit RMSS ensembles  is also available on CRAN in the \texttt{R} package \texttt{RMSS} \citep{RMSS_package}, which generates RBSS if $ G = 1 $. The source code of \texttt{RMSS} is written in \texttt{C++}, and multithreading via OpenMP \citep{chandra2001parallel} is available in the package to further speed up computations.
	A compressed archive containing a detailed \texttt{README} file, all \texttt{R} scripts and data files necessary to reproduce the numerical results is on Zenodo (\url{https://doi.org/10.5281/zenodo.16915351}).
	
	\section{Simulations} \label{sec:simulation}
	
	In this section, we investigate the performance of RMSS against robust and sparse methods as well as blackbox ensemble methods in an extensive simulation study where the data is contaminated %with leverage points, i.e., samples contaminated 
	in both the predictor space and the response. We also use a block correlation structure between predictor variables to mimic as closely as possible the behavior of many modern datasets \citep{zhang2012sources}.

	\subsection{Simulation of Uncontaminated Data}
	
	In each setting of our simulation study, we generate the uncontaminated data from the linear
	model %(\ref{eq:true_model}),
	
	\begin{equation*}
		y_{i} = \mu + \mathbf{x}_{i}^{T} \boldsymbol{\beta}_{0} + \sigma \epsilon_{i},
		\quad 1\leq i \leq n,
	\end{equation*}
	with independent and identically distributed $\bx_i \sim N_p(\boldsymbol{0}, \boldsymbol{\Sigma})$ and $\epsilon_i \sim N(0,1)$, sample size $n= 50$ and the number of predictors  $p = 500$. Alternative choices for  $n$ and $p$ where $ p \gg n $ lead to similar conclusions  and are omitted for conciseness.
	
	%$\mat x_i \sim N_p(\mat 0, \mat{\Sigma})$ 
	%To generate disjoint blocks of correlated of active predictors (i.e., with nonzero coefficients) we use a within-block correlation $\rho_1 = 0.8$ and  a between-block correlation $ \rho_2 = 0.2 $. The non-active predictors are independent and uncorrelated with the active ones. We set the number of blocks of correlated predictors so that each block is comprised of 25 predictors. Based on our preliminary numerical experiments, alternative choices for  $n$ and $p$ where $ p \gg n $ lead to similar conclusions.
	%For each $p$, we consider the proportion of active variables  equal to $\zeta \in \{0.1, 0.2, 0.4\}$. For simplicity, the intercept is set to $\mu =0$. The coefficients of the active variables, $\{\beta_j: \beta_j\neq 0, 1 \leq j \leq p\}$, are randomly generated from the random variable $ (-1)^Z  \times U $, where $Z$ is Bernoulli distributed with parameter $ 0.7 $ and $ U $ is uniformly distributed on the interval $\left(0, 5\right)$.
	%The noise parameter $\sigma$ is computed based on the desired signal to noise ratio,  $\text{SNR} = {\bbet_{0}^{\prime} \boldsymbol{\Sigma} \bbet_{0}}/{\sigma^2}$.We consider SNRs of 0.5 (low signal), 1 (moderate signal), 2 (high signal), which correspond to proportions of variance explained $\text{PVE}= \text{SNR} / (\text{SNR} + 1) $  of $ 33.3\% $, $ 50\% $ and $ 66.7\% $, respectively.
	
	We examine models with different proportions $\zeta$ of active predictors, with $\zeta \in \{0.1, 0.2, 0.4\}$. Without loss of generality, we construct the correlation matrix of the active predictors as a block matrix, with each block corresponding to 25 predictors, a within-block correlation $\rho_1 = 0.8$ and a between-block correlation $ \rho_2 = 0.2 $. The non-active predictors are independent and uncorrelated with the active ones. 
	
	For simplicity, the intercept is set to $\mu =0$. The coefficients of the active variables, $\{\beta_j: \beta_j\neq 0, 1 \leq j \leq p\}$, are randomly generated from the random variable $ (-1)^Z  \times U $, where $Z$ is Bernoulli distributed with parameter $ 0.7 $ and $ U $ is uniformly distributed on the interval $\left(0, 5\right)$. 
	
	The noise parameter $\sigma$ is computed based on the desired signal to noise ratio,  $\text{SNR} = {\bbet_{0}^{\prime} \boldsymbol{\Sigma} \bbet_{0}}/{\sigma^2}$. We consider SNRs of 0.5 (low signal), 1 (moderate signal), 2 (high signal), which correspond to proportions of variance explained $\text{PVE}= \text{SNR} / (\text{SNR} + 1) $  of $ 33.3\% $, $ 50\% $ and $ 66.7\% $, respectively.
	
	\subsection{Data Contamination}
	
	We contaminate the first $ m = \lfloor \alpha n  \rfloor $ samples $ (\bx_i, y_i) $ according to the model proposed in \cite{maronna2011robust}. The regression outliers  are introduced by replacing the predictors $ \bx_i $ with 
	\begin{align*}
		\mathbf{\tilde{x}_i} = \Theta_i + \frac{\text{k}_{\text{lev}}}{\sqrt{\mathbf{a}^T \mathbf{\Sigma}^{-1}\mathbf{a}}}\mathbf{a}, \quad 1 \leq i \leq m,
	\end{align*}
	where $ \Theta_i \sim \mathcal{N}(\mathbf{0_p}, 0.01 \times \mathbf{I_p}) $ and $ \mathbf{a} = \mathbf{\tilde{a}} - (1/p) \mathbf{\tilde{a}}^T\mathbf{1_p} $ where 
	$\mathbf{I_p}$ is the $p$-dimensional identity matrix, $ \mathbf{0_p} = (0, \dots, 0)^T \in \mathbb{R}^p $, $ \mathbf{1_p} = (1, \dots, 1)^T \in \mathbb{R}^p $, and the entries of $\tilde{a}_j$ of $ \mathbf{\tilde{a}} $ follow a uniform distribution on the interval $ (-1, 1) $, $ 1 \leq j \leq p $. The parameter $ \text{k}_{\text{lev}} $ controls the distance in the direction most influential for the estimator.
	
	We also contaminate the observation in the response by altering the regression coefficient
	\begin{align*}
		\tilde{y}_i = \mathbf{\tilde{x}_i}^T \tbbet, \quad \tilde{\beta}_j = \begin{cases}
			\beta_j(1+\text{k}_{\text{slo}}), &\beta_j \neq 0, \\
			\text{k}_{\text{slo}} \left\lVert \bbet \right\rVert_{\infty}, &\text{otherwise},
		\end{cases}
		\quad 1 \leq i \leq m.
	\end{align*} 
	
	The parameters $k_{\text{lev}}$ and $k_{\text{slo}}$ control the position of contaminated observations. Preliminary experiments showed that estimator performance was nearly unchanged for any $k_{\text{lev}}>1$, so we fixed $k_{\text{lev}}=2$. In contrast, the position of vertical outliers had a much stronger impact: non-robust estimators degraded markedly for any $k_{\text{slo}}\ge100$, so we fixed $k_{\text{slo}}=100$. We consider contamination proportions $\alpha=0$ (none), $\alpha=0.15$ (moderate), and $\alpha=0.3$ (high).
	
	\subsection{Methods}
	
	Our simulation study compares the prediction and variable selection accuracy of eight methods. All computations were carried out in \texttt{R} using the implementations listed below	\begin{enumerate}
		\item[1.] Elastic Net \citep[\textbf{EN},][]{zou2005regularization},  with \texttt{glmnet} package \citep{glmnet}. 
		\item[2.] Adaptive \textbf{PENSE} \citep{kepplinger2023robust},  with \texttt{pense} package \citep{pense_package}.
		\item[3.] EN Penalized Huber \citep[\textbf{HuberEN},][]{yi2017semismooth},  with \texttt{hqreg} package \citep{hqreg_package}.
		\item[4.] Sparse LTS \citep[\textbf{SparseLTS},][]{alfons2013sparse}, with \texttt{robustHD} package \citep{robustHD_package}.
		\item[5.] Robust Best Subset Selection \citep[\textbf{RBSS},][]{thompson2022robust},  with \texttt{RMSS} package.
		\item[6.] Robust Multi-Model Subset Selection (\textbf{RMSS}) with $ G = 10 $ models, proposed in this paper,  with \texttt{RMSS} package.
		\item[7.] {Random GLM} \citep[\textbf{RGLM},][]{random_glm_paper},  with \texttt{RMSS} package.
		\item[8.] {Random Forest} \citep[\textbf{RF},][]{RF}, with \texttt{randomForest} package \citep{randomForest}.
		
	\end{enumerate}
	
	To reduce computational cost in our extensive simulation study, we used modest tuning grids for RMSS. For the sparsity parameter, we used the candidate grid $\mathcal{T} = \{15, 20, 25\}$. The robustness parameter $h$ was selected via CV from the grid $\mathcal{H} = \{(1 - (\alpha + 0.1))n, (1 - (\alpha + 0.05))n, (1 - \alpha)n\}$, where $\alpha$ is the true contamination level. We used the default grid $\mathcal{U}=\{1, \dots, G\}$ with $G=10$ for the diversity parameter. RBSS was computed concurrently by evaluating the $u=G$ case (see Proposition \ref{prop:rmss_special_cases}). While more refined grids could yield further improvements, these settings proved sufficient for RMSS to be highly competitive. We used 5-fold CV for the robust methods and 10-fold CV for EN; additional details are in Appendix G.
	
	\subsection{Performance Measures}
	
	For each simulation configuration (i.e., each combination of sparsity $\zeta$, SNR, and contamination $\alpha$), we generated $N=50$ training sets and a single, large, uncontaminated test set of size 2,000. For each replication, we fit the methods on the training data and evaluated performance on the test set.

	We report three performance metrics. The Mean Squared Prediction Error (MSPE) is reported relative to the irreducible error variance, $\sigma^2$, making the optimal value 1. For variable selection, we compute Recall (RC) and Precision (PR), defined as:
	\begin{align*}
		\text{RC} = \frac{\sum_{j=1}^p\mathbb{I}(\beta_j\neq 0, \hat{\beta}_j\neq0)}{\sum_{j=1}^p\mathbb{I}(\beta_j\neq 0)}, \quad \text{PR} = \frac{\sum_{j=1}^p\mathbb{I}(\beta_j\neq 0, \hat{\beta}_j\neq0)}{\sum_{j=1}^p\mathbb{I}(\hat{\beta}_j\neq0)},
	\end{align*} 
	where $\bbet$ and $\boldsymbol{\hat{\beta}}$ are the true and estimated coefficients, respectively. For RMSS, we use the averaged ensemble fit. We do not report RC and PR for RGLM and RF, as their dense nature makes these metrics uninformative. For both RC and PR, higher values are better.

	\subsection{Results}
	
	In Table \ref{tab:ranks_mspe}, we report the average and lowest MSPE rank of the eight methods over the nine SNR and sparsity level combinations for each contamination proportion. The top two performances in each column are highlighted in bold.

\begin{figure}[ht!]
		\centering
		\includegraphics[width=17.5cm]{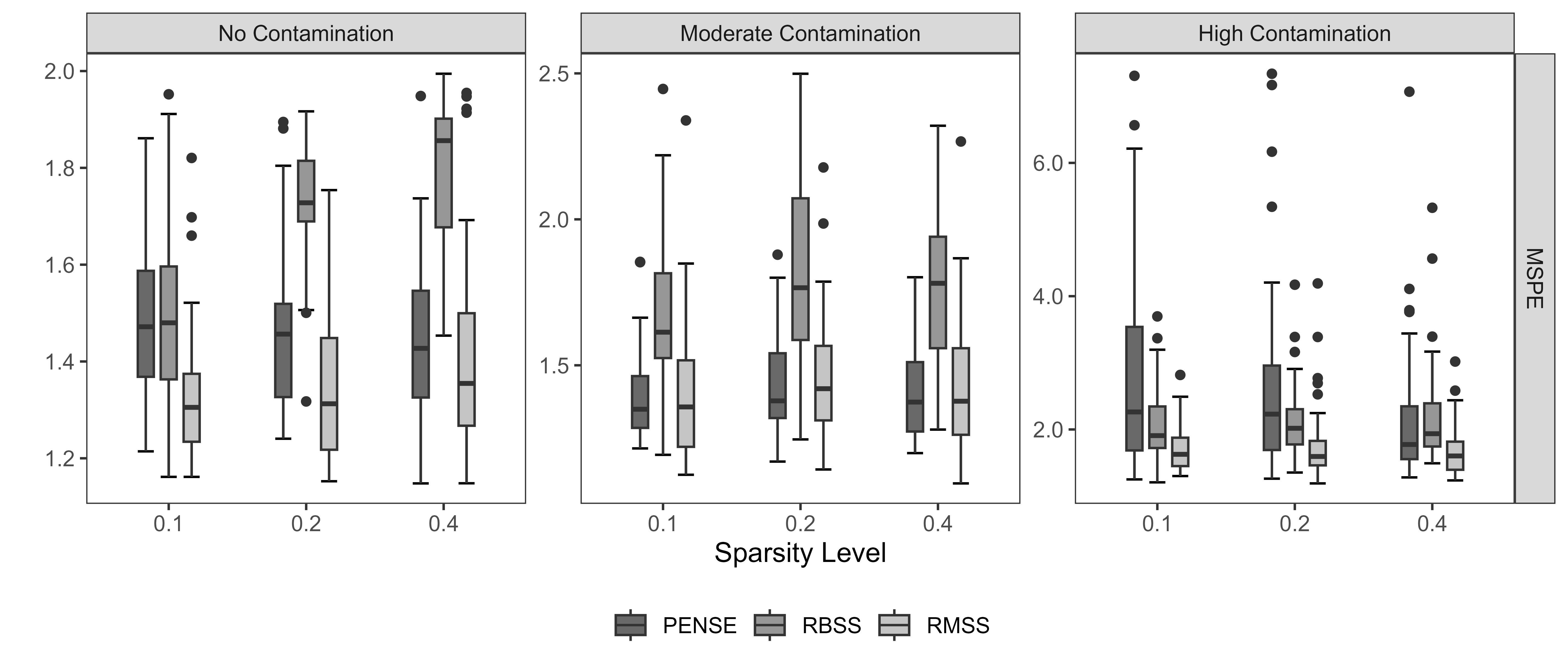}
		\caption{MSPE of PENSE, RBSS and RMSS over $ N=50 $ random training sets over different contamination and sparsity levels for $ \text{SNR} = 1 $.}
		\label{fig:Simulation_Plot_MSPE}
	\end{figure}

	In the uncontaminated case ($\alpha=0$), RGLM achieved the best overall rank, followed by RMSS and RF. RMSS consistently outperformed the non-robust EN, demonstrating its ability to adapt well to clean data. For moderate contamination ($\alpha=0.15$), PENSE and RMSS were the clear top performers, ranking first and second, respectively, across every simulation setting. As shown in Figure \ref{fig:Simulation_Plot_MSPE}, their prediction performances were very similar under this scenario. However, in the high contamination case ($\alpha=0.30$), PENSE's performance deteriorated while RMSS achieved the top rank in every configuration. The non-robust ensemble methods' predictive performance degraded completely under both moderate and high contamination.

	To better contextualize these summary results, Figure \ref{fig:Simulation_Plot_MSPE} plots the MSPE of the top robust competing methods (PENSE, RBSS, and RMSS) from the $N=50$ random training sets for a moderate SNR. It is evident that while PENSE and RMSS perform very similarly at moderate contamination levels, RMSS significantly outperforms both PENSE and RBSS when the contamination level is high. This conclusion holds across all three SNR levels considered in our study.
	
	\begin{table}[ht!]
		\centering
		\caption{Average and lowest MSPE rank of the eight methods over the nine  SNR and  sparsity level combinations for  each contamination proportion.  \label{tab:ranks_mspe}} 
		\extrarowsep =2pt
		\begin{tabu}{llcccccc}
			\toprule 
			&& \multicolumn{6}{c}{\textbf{MSPE Rank}} \\ 
			\cmidrule(lr){3-8}
			&& \multicolumn{2}{c}{$\mathbf{\boldsymbol{\alpha}=0}$} & \multicolumn{2}{c}{$\mathbf{\boldsymbol{\alpha}=0.15}$} & \multicolumn{2}{c}{$ \mathbf{\boldsymbol{\alpha}=0.3} $} \\ 
			\cmidrule(lr){1-1}  \cmidrule(lr){3-4} \cmidrule(lr){5-6} \cmidrule(lr){7-8}
			\textbf{Method} && \textbf{Avg} & \textbf{Low}  & \textbf{Avg} & \textbf{Low}  & \textbf{Avg} & \textbf{Low} 
			\\ 
			\cmidrule(lr){1-1}  \cmidrule(lr){3-4} \cmidrule(lr){5-6} \cmidrule(lr){7-8}
			\addlinespace[0.25cm]
			%	EN && 3.2 & 4 & {6.0} & 6 & 4.7 & 5  \\ 
			%	PENSE && 2.4 & 5 & \textbf{1.2} & \textbf{2} & 4.1 & 5  \\ 
			%	HuberEN && 3.2 & 4 & 4.4 & 5 & 2.8 & 4  \\ 
			%	SparseLTS && 5.7 & 6 & 3.3 & 4 & 6.0 & 6  \\ 
			%	RBSS && 5.1 & {6} & 4.2 & 5 & 2.4 & 4 \\ 
			%	RMSS && \textbf{1.3} & \textbf{2} & 1.8 & \textbf{2} & \textbf{1.0} & \textbf{1} \\ 
			EN         && 5.3       & 6         & 5.9       & 6         & 4.6       & {5}         \\ 
			PENSE   && 4.2       & 7         & \textbf{1.0} & \textbf{1}          & 3.9       & {5}         \\ 
			HuberEN  && 5.0       & 6         & 4.4       & 5         & \textbf{2.4}      & \textbf{4}         \\ 
			SparseLTS  && 7.9       & 8         & 3.3       & 4         & 7.0       & 8         \\ 
			RBSS       && 6.9       & 8         & 4.3       & 6         & {3.1}       & {5}         \\ 
			RMSS       && \textbf{2.7}       & \textbf{4}         & \textbf{2.0}       & \textbf{2}         & \textbf{1.0} & \textbf{1}        \\ 
			RGLM   && \textbf{1.1}       & \textbf{2}         & 8.0       & 8         & 7.7       & 8         \\ 
			RF         && 2.9       & 5         & 7.0       & 7         & 6.3       & 7         \\ 
			\bottomrule
		\end{tabu}
	\end{table}

	Table \ref{tab:ranks_selection} summarizes the variable selection performance, reporting the average Recall (RC) and Precision (PR) ranks for the six relevant methods across all simulation settings. RMSS consistently achieved the best average RC rank across all contamination levels, with PENSE as the next best performer. In terms of PR, RBSS performed best, followed by RMSS in the contaminated scenarios.

\begin{figure}[ht!]
		\centering
		\includegraphics[width=17.5cm]{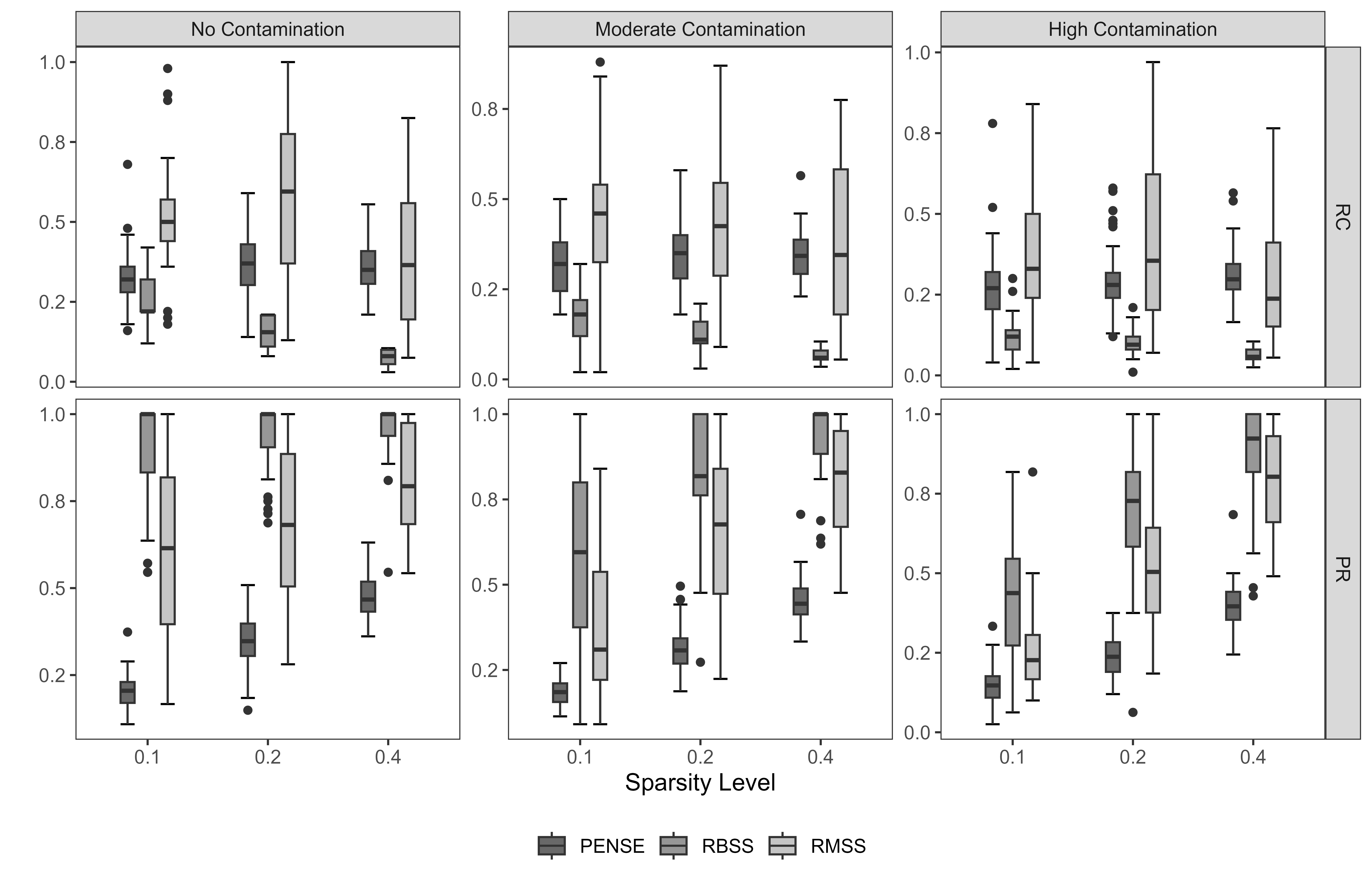}
		\caption{RC and PR of PENSE, RBSS and RMSS over $ N=50 $ random training sets over different contamination and sparsity levels for $ \text{SNR} = 1 $.}
		\label{fig:Simulation_Plot_RCPR}
	\end{figure}

	Figure \ref{fig:Simulation_Plot_RCPR} provides a more detailed view for a moderate SNR. RMSS generally outperformed its competitors in RC, especially for lower-sparsity models ($\zeta=0.1, 0.2$), a result that improved further in experiments with more than $G=10$ models. While RBSS consistently achieved the highest PR, this came at the cost of poor RC across all settings. In contrast, RMSS provided the best balance, combining high RC with high PR across all configurations of our simulation study. This demonstrates its ability to reliably identify true predictors while controlling for false positives, even in complex data with block correlation and regression outliers.
	
	\begin{table}[ht!]
		\centering
		\caption{Average and lowest RC and PR rank of the six methods over the nine SNR and  sparsity level combinations for  each contamination proportion.  \label{tab:ranks_selection}} 
		\resizebox{\textwidth}{!}{
			\extrarowsep =2pt
			\begin{tabu}{llccccccccccccc}
				\toprule 
				& &\multicolumn{6}{c}{\textbf{RC Rank}} && \multicolumn{6}{c}{\textbf{PR Rank}} \\ 
				\cmidrule(lr){3-8} \cmidrule(lr){10-15}
				&& \multicolumn{2}{c}{$\mathbf{\boldsymbol{\alpha}=0}$} & \multicolumn{2}{c}{$\mathbf{\boldsymbol{\alpha}=0.15}$} & \multicolumn{2}{c}{$ \mathbf{\boldsymbol{\alpha}=0.3} $} && \multicolumn{2}{c}{$\mathbf{\boldsymbol{\alpha}=0}$} & \multicolumn{2}{c}{$\mathbf{\boldsymbol{\alpha}=0.15}$} & \multicolumn{2}{c}{$ \mathbf{\boldsymbol{\alpha}=0.3} $} \\ 
				\cmidrule(lr){1-1}  \cmidrule(lr){3-4} \cmidrule(lr){5-6} \cmidrule(lr){7-8} \cmidrule(lr){10-11}  \cmidrule(lr){12-13} \cmidrule(lr){14-15}
				\textbf{Method} && \textbf{Avg} & \textbf{Low}  & \textbf{Avg} & \textbf{Low}  & \textbf{Avg} & \textbf{Low} && \textbf{Avg} & \textbf{Low}  & \textbf{Avg} & \textbf{Low}  & \textbf{Avg} & \textbf{Low} 
				\\ 
				\cmidrule(lr){1-1}  \cmidrule(lr){3-4} \cmidrule(lr){5-6} \cmidrule(lr){7-8} \cmidrule(lr){10-11}  \cmidrule(lr){12-13} \cmidrule(lr){14-15}
				\addlinespace[0.25cm]
				EN && 5.6 & 6 & {5.7} & 6 & 5.3 & {6} & & \textbf{2.2} & \textbf{4} & 5.7& 6 & 5.3 & 6\\ 
				PENSE && 2.6 & \textbf{3} & \textbf{1.9} & \textbf{2} & \textbf{1.7} & \textbf{2} && {5.0} & 5 & 3.0& 3 & 3.0 & 3 \\ 
				HuberEN && \textbf{2.1} & {4} & 5.1 & 6 & 5.7 & 6  && 3.7 & \textbf{4} & 5.3& 6 & {5.7} & {6}\\ 
				SparseLTS && 5.4 & 6 & 4.1 & 5 & 4.0 & 4& & 6.0 & 6 & 4.0& {4} & 4.0 & 4 \\ 
				RBSS && {3.9} & 4 & {3.1} & \textbf{4} & {3.0} & \textbf{3} && \textbf{1.1} & \textbf{2} & \textbf{1.0} & \textbf{1} & \textbf{1.0} & \textbf{1}\\ 
				RMSS && \textbf{1.4} & \textbf{2} & \textbf{1.1} & \textbf{2} & \textbf{1.3} & \textbf{2}& & {3.0} & \textbf{4} & {\textbf{2.0}}& {\textbf{2}} & {\textbf{2.0}} & {\textbf{2}}\\ 	
				\bottomrule
		\end{tabu}}
	\end{table}
	
	\subsection{Computing Times}
	
	Table \ref{tab:cpu} reports the average computation time for each method across all simulation configurations. Notably, our RMSS implementation, which simultaneously generates $G=10$ robust models and performs a 3D cross-validation, achieved a lower average computing time than PENSE. This is particularly efficient given that PENSE generates only a single model and performs CV over one parameter. The reported time for RMSS also includes the computation of RBSS, as it is generated concurrently by setting $u=G$.

	\begin{table}[ht!]
		\centering
		\caption{\label{tab:cpu}Computation time of \texttt{R} function calls for the methods in CPU seconds. CPU seconds are on a 2.1 GHz Intel Xeon Platinum 8468 processor in a machine running  CentOS Linux 7.9 with 32 GB of RAM.} 
		\extrarowsep=2pt
		\begin{tabu}{lcccccc}
			\toprule
			%\tabucline[1.5pt]{}
			\textbf{Method}    && EN    & PENSE     & HuberEN     & SparseLTS   & RMSS       \\ \hline
			\rule{0pt}{4ex}\textbf{Time} && 0.2 & 329.2 & 0.1 & 131.8 & 137.7  \\
			\bottomrule
			%\tabucline[1.5pt]{}
		\end{tabu}
	\end{table}
	
\begin{figure}[ht!]
		\centering
		\includegraphics[width=15.5cm]{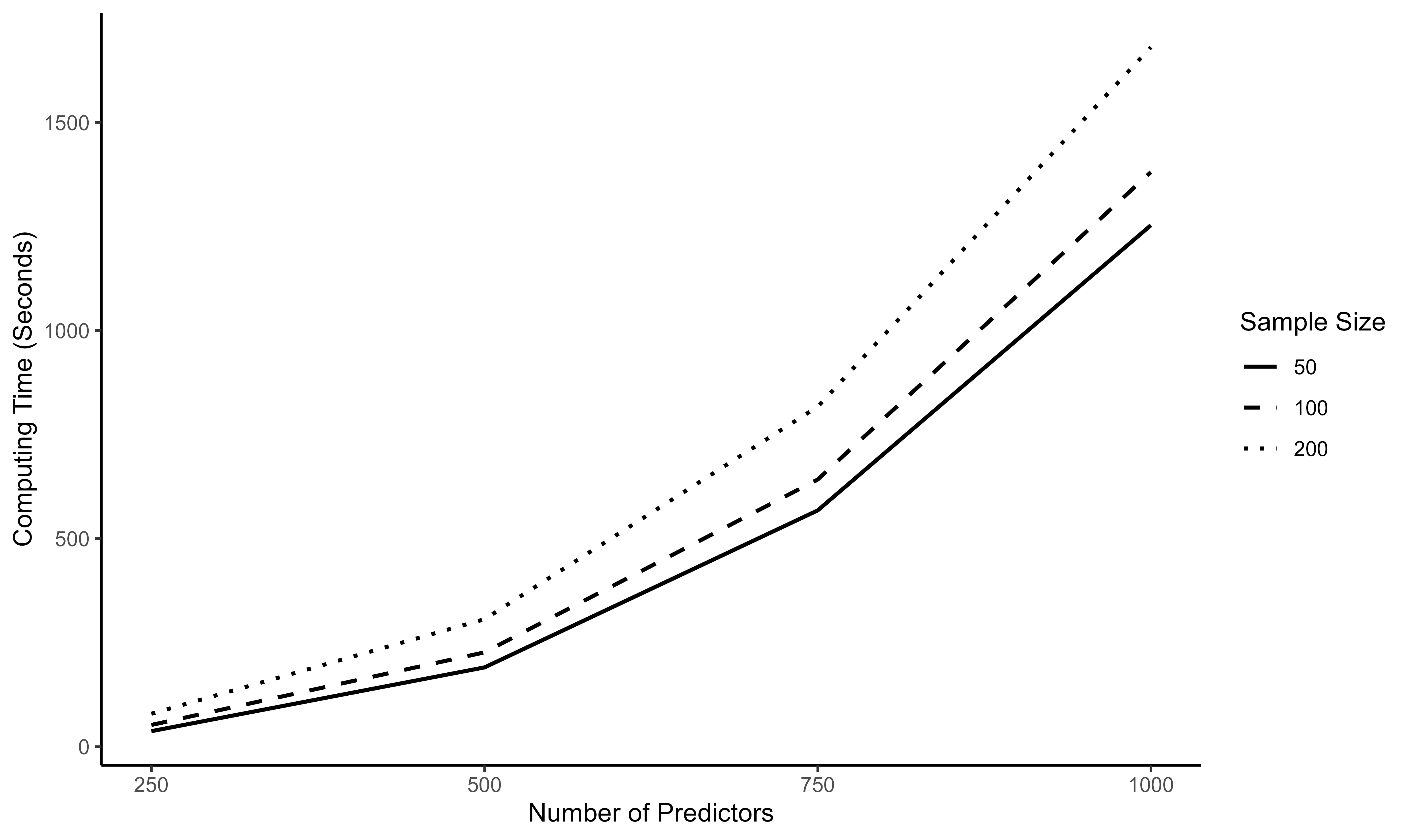}
		\caption{Computing time versus number of predictors ($p = 250, 500, 750, \text{1,000}$) for different sample sizes ($n = 50, 100, 200$). Results show mean elapsed time across replications.}
		\label{fig:Computing_Plot}
	\end{figure}

	To assess scalability for larger datasets, we examined how the computing time of RMSS varies with sample size $n$ and number of predictors $p$. We used a challenging simulation scenario (low signal, high contamination, high proportion of true predictors) to provide a realistic performance benchmark. Figure \ref{fig:Computing_Plot} displays the run time as a function of $p \in \{250, \dots, 1000\}$ for different sample sizes $n \in \{50, 100, 200\}$. Results show mean elapsed time across replications. As previously noted for other estimators based on mixed-interger optimization \citep{thompson2022robust, bertsimas2016best}, RMSS remains computationally tractable up to dimensions as large as p = 1000, despite the expected increase in runtime with more predictors. 	
	
	The reported times do not include the optional neighborhood search strategy described in Appendix F since we found that it significantly increases the computational cost with only marginal gains in prediction accuracy. Overall, this analysis offers a guidance on the computational times expected for datasets of varying dimensions and highlights the practical scalability of our algorithm.

	\section{Contamination of Bioinformatics and Cheminformatics Data} \label{sec:artificial_contamination}
	
	In biomedical sciences new deoxyribonucleic acid (DNA) microarray and  ribonucleic acid (RNA) sequencing technologies allow for an increase in the type and volume  of the genomics data collected \citep[e.g.,][]{byron2016translating}. In chemistry, innovative microscopic technologies allow for the collection of data on the composition of chemical compounds and molecules. Outliers are not rare in these datasets. For example, in genomics, many datasets contain atypical observations obtained from samples with poor measurement quality or incorrect reads \citep[e.g.,][]{sangiovanni2019trash}.  In cheminformatics the topic of high-dimensionality and robustness has gained a lot of attention in recent years due to the emerging fields of computer-aided drug design and computational toxicology among others \citep[see e.g.][]{basak2022big}.
	
	In this section we artificially contaminate real bioinformatic and cheminformatic datasets to evaluate the performance of RMSS and the other methods in situations that mimic real applications. We also show that RMSS can uncover some predictors that may be relevant to predict the outcome of interest but that may not be picked up by single-model sparse robust methods.	
	
	\subsection{Bioinformatics Data}
	
	We first analyze data from a study on Bardet-Biedl syndrome (BBS) by \cite{scheetz2006regulation}, using a dataset from the \texttt{R} package \texttt{abess} \citep{zhu2022abess}. The task is to predict the expression of the gene TRIM32 using $p=500$ other genes from 120 mammalian-eye tissue samples. We randomly split the data 50 times into training ($n=50$) and test sets ($n=70$). We then contaminated 25\% of samples in each training set by replacing the TRIM32 expression and 100 predictor genes with large, atypical values.

	The relative MSPE and standard deviation (SD) in Table \ref{tab:bioinformatics} show that RMSS achieved the best performance with the lowest variability. PENSE, the closest competitor, had an MSPE 7\% higher than RMSS. As expected, the non-robust EN method failed completely, indicating the RMSS ensemble effectively handles data contamination.
	
	\begin{table}[ht!]
		\centering
		\caption{MSPE and SD relative to the best performance for the artificially contaminated BBS dataset ($p=500$). \label{tab:bioinformatics}} 
		\extrarowsep=2pt
		\begin{tabu}{lccccccc}
			\toprule
			\textbf{Method}    && EN    & PENSE     & HuberEN     & SparseLTS & RBSS  & RMSS       \\ \hline
			\rule{0pt}{4ex}\textbf{MSPE} && $ >25 $ & 1.07 & 1.17 & {1.09} & 1.74 & \textbf{1.00}  \\
			\textbf{SD} && $ > 75 $ & 1.15 & 1.22 & {1.37} & 2.18 & \textbf{1.00}  \\
			\bottomrule
		\end{tabu}
	\end{table}
	
	Beyond its predictive accuracy, RMSS provides a more comprehensive view of relevant predictors. This is valuable in high-dimensional settings due to the ``multiplicity of good models" phenomenon \citep{mccullagh1989monographs}. Single-model methods may overlook important genes; indeed, on the BBS dataset, both PENSE and SparseLTS exhibited unstable variable selection, with no gene selected in over 50\% of the runs. In contrast, RMSS identified 14 genes with over 50\% selection frequency. It not only captured predictors favored by others but also consistently identified genes they missed, such as AAK1 (selected in 68\% of RMSS runs vs. only 2\% for PENSE and SparseLTS).

	To evaluate performance in a higher dimensional setting, we conducted an additional analysis using the original microarray data from the GSE5680 dataset \citep{barrett2005ncbi}. After standard preprocessing, we created a feature set of $p=\text{1,275}$ predictors by selecting the 50 genes most correlated with TRIM32 and including all their pairwise interactions. This biologically-motivated approach creates a challenging $p/n \approx 25$ scenario. We applied the same 25\% contamination framework to this data.
	
	The results in Table \ref{tab:bioinformatics_hd} show that RMSS continues to achieve excellent prediction performance. Despite the significant increase in dimensionality, RMSS maintains its robustness advantage over competing methods, confirming its applicability to modern genomic datasets where predictors greatly exceed the sample size.
	
	\begin{table}[ht!]
		\centering
		\caption{MSPE and SD relative to the best performance for the artificially contaminated BBS dataset with interactions ($p=1,275$). \label{tab:bioinformatics_hd}} 
		\extrarowsep=2pt
		\begin{tabu}{lccccccc}
			\toprule
			\textbf{Method}    && EN    & PENSE     & HuberEN     & SparseLTS & RBSS  & RMSS       \\ \hline
			\rule{0pt}{4ex}\textbf{MSPE} && $> 50$ & 1.08 & 1.94 & 1.21 & 1.35 & \textbf{1.00}  \\
			\textbf{SD} && $>50$ & 3.93 & 2.84 & 1.96 & 2.06 & \textbf{1.00}  \\
			\bottomrule
		\end{tabu}
	\end{table}
	
	\subsection{Cheminformatics Data}
	
	We next analyze the glass dataset from \cite{lemberge2000quantitative}, where the goal is to predict the concentration of the chemical compound Na2O from $p=486$ frequency measurements obtained via electron probe X-ray microanalysis (EPXMA). Using 50 random splits, we created training sets ($n=50$) and test sets ($n=130$). We applied the same 25\% contamination scheme to the training sets and evaluated performance on the uncontaminated test sets.

	As shown in Table \ref{tab:cheminformatics}, which reports relative MSPE and its standard deviation, RMSS again achieved the best predictive performance by a large margin. PENSE, the next best method, had an MSPE 32\% larger than RMSS. The individual models within the RMSS ensemble also performed well, achieving an MSPE comparable to the single-model RBSS solution. This strong predictive performance was observed for other chemical compounds in the full dataset as well.
	
	\begin{table}[ht!]
		\centering
		\caption{MSPE and SD relative to the best performance for the artificially contaminated glass cheminformatics dataset. SparseLTS was omitted as it consistently returned an empty model. \label{tab:cheminformatics}} 
		\extrarowsep=2pt
		\begin{tabu}{lcccccc}
			\toprule
			\textbf{Method}    && EN    & PENSE     & HuberEN      & RBSS  & RMSS       \\ \hline
			\rule{0pt}{4ex}\textbf{MSPE} && $ >100 $ & 1.32 & 8.46  & 2.45 & \textbf{1.00}  \\
			\textbf{SD} && $ > 75 $ & 1.21 & 1.62  & 2.34 & \textbf{1.00}  \\
			\bottomrule
		\end{tabu}
	\end{table}
	
	The superior variable selection of RMSS was also evident. The ensemble consistently identified important frequency measurements that PENSE selected unreliably. This is likely because PENSE's elastic net penalty tends to select only one predictor from a correlated group, whereas RMSS's ensemble structure can accommodate such predictors by assigning them to different models. This was clear in the results: the top predictor for RMSS (86\% selection rate) was chosen by PENSE in only 12\% of runs. Further, a feature PENSE selected in just 26\% of runs was captured far more consistently by the RMSS ensemble (78\% selection rate). This highlights the ability of RMSS to provide a more robust and complete feature set, particularly under multicollinearity.

	\section{Summary and Future Works} \label{sec:summary_future}
	
	In this article, we introduce RMSS, a data-driven method to build an ensemble of sparse and robust models to predict a response of interest and select important predictors from moderately high-dimensional datasets possibly containing outlying observations. To the best of our knowledge, this is the first method proposed for this aim. 
	The levels of sparsity, diversity and robustness of each model are driven directly by the data based on a CV criterion.
	We established the finite-sample breakdown point of RBSS and RMSS. To bypass the NP-hard computational complexity of RMSS, we  developed a tailored computing algorithm with a local convergence property by leveraging recent developments in the $ \ell_0 $-optimization literature. Our extensive numerical experiments on synthetic and real data demonstrate the excellent performance of RMSS relative to state-of-the-art sparse and robust methods in moderately high-dimensional prediction tasks when the data is also contaminated. We also showed how RMSS can potentially uncover important predictor variables that may be discarded by single-model sparse and robust methods.
	
	Since RMSS can potentially uncover predictor variables that are not picked up by single-model methods, the addition of interaction terms may potentially further increase the competitive advantage of RMSS over single-model sparse and robust methods. For example, in the -omics sciences where interactions between genes or proteins may drive the outcome of interest. The empirical performance of RMSS can  be improved further by considering alternative ways to combine the models in the ensembles other than the simple model average we used in this article. Our work can be extended by considering other robust loss functions to build sparse robust models.
	
	With the growing emphasis on interpretable statistical and machine learning algorithms in the literature and in real data applications, our proposal will potentially pave the way for the development of other robust ensemble methods. A potential bottleneck in this area of research is the high computational cost of such methods, thus new optimization tools will be needed to render such ensemble methods feasible in practice.

	\section*{Appendix A: Special Cases of RMSS}
	
	In the multi-part proof of Proposition 1 below, we demonstrate how RMSS generalizes several well-known estimators (MSS, RBSS, BSS, and LTS) under specific settings of its tuning parameters $t, u,$ and $h$.
	
	The proof for each part follows by analyzing how the constraints and objective function of the RMSS problem  simplify under the specified conditions. Let $\mathcal{I} = \{1, \dots, n\}$.
	
	\subsection*{If $h=n$, RMSS is Equivalent to MSS}
	
	When the trimming parameter $h$ is set to $n$, the constraint $|\mathcal{I}^g| \geq h$ for each model $g$ becomes $|\mathcal{I}^g| \geq n$. Since each $\mathcal{I}^g$ is a subset of $\mathcal{I}$ (where $|\mathcal{I}|=n$), the only way to satisfy this constraint is for $\mathcal{I}^g = \mathcal{I}$ for all $g$, $1 \leq g \leq G$. This effectively removes the optimization over the observation subsets $\mathcal{I}^g$ and fixes the loss function to be the standard sum of squared residuals over all data points. The RMSS problem simplifies to:
	\begin{align*} 
		\min_{\bbet^1, \dots, \bbet^G \in \mathbb{R}^p} \sum_{g=1}^{G} \sum_{i=1}^{n} \left(y_i - \bx_i^T \bbet^g\right)^2 \quad \text{subject to} \quad \begin{cases}
			\left\lVert\bbet^g\right\rVert_0 \leq t, \, &1 \leq g \leq G, \\
			\left\lVert\bbet_{j\cdot}\right\rVert_0 \leq u, \, & 1 \leq j \leq p.
		\end{cases}
	\end{align*}
	This is precisely the formulation for the Multi-Model Subset Selection (MSS) estimator. \hfill $\square$
	
	\subsection*{\bf If $u=G$, RMSS is Equivalent to Solving $G$ Independent RBSS Problems}
	
	When the diversity parameter $u$ is set to $G$, the diversity constraint $\left\lVert\bbet_{j\cdot}\right\rVert_0 \leq u$ becomes $\left\lVert\bbet_{j\cdot}\right\rVert_0 \leq G$. Since $\bbet_{j\cdot} \in \mathbb{R}^G$, its $\ell_0$-norm can be at most $G$. This constraint is therefore always satisfied and becomes non-binding (inactive). With the coupling constraint removed, the RMSS objective function, which is a sum over the $G$ models, becomes separable. The optimization can be decomposed as follows:
	\begin{align*} 
		&\min_{\substack{\bbet^1, \dots, \bbet^G \in \mathbb{R}^p \\ \mathcal{I}^1, \dots, \mathcal{I}^G \subseteq \mathcal{I}}} \sum_{g=1}^{G} \sum_{i \in \mathcal{I}^g} \left(y_i - \bx_i^T \bbet^g\right)^2 \quad \text{subject to} \quad \begin{cases}
			\left\lVert\bbet^g\right\rVert_0 \leq t, \, &1 \leq g \leq G, \\
			|\mathcal{I}^g| \geq h, \, & 1 \leq g \leq G.
		\end{cases}\\[1em]
		=& \sum_{g=1}^{G} \; \left[ \min_{\substack{\bbet^g \in \mathbb{R}^p \\ \mathcal{I}^g \subseteq \mathcal{I}}} \; \sum_{i \in \mathcal{I}^g} \left(y_i - \bx_i^T \bbet^g\right)^2 \quad \text{subject to} \quad \begin{cases}
			\left\lVert\bbet^g\right\rVert_0 \leq t, \\
			|\mathcal{I}^g| \geq h.
		\end{cases} \right]
	\end{align*}
	Each term in the sum is an identical, independent optimization problem. This subproblem is precisely the formulation for Robust Best Subset Selection (RBSS). Therefore, the solution for each model, $\hbbet^g$, is an optimal solution to the RBSS problem. \hfill $\square$
	
	\subsection*{If $u=G$ and $h=n$, RMSS is Equivalent to BSS}
	
	This follows directly from combining the results of (I) and (II). From (II), we know that when $u=G$, the RMSS problem decomposes into solving $G$ independent RBSS problems. If we additionally set $h=n$, then as shown in (I), the robustness constraint $|\mathcal{I}^{\prime}| \geq n$ forces the trimming set to be the full set of observations, $\mathcal{I}^{\prime} = \mathcal{I}$. The RBSS subproblem thus simplifies to:
	\begin{align*}
		\min_{\bbet \in \mathbb{R}^p} \; \sum_{i=1}^{n} \left(y_i - \bx_i^T \bbet\right)^2 \quad \text{subject to} \quad \left\lVert\bbet\right\rVert_0 \leq t.
	\end{align*}
	This is the formulation for Best Subset Selection (BSS). \hfill $\square$

	\subsection*{If $u=G$ and $t=p < n-1$, RMSS is Equivalent to LTS}
	
	This also follows from the result of (II). When $u=G$, RMSS reduces to solving $G$ independent RBSS problems. If we additionally set the sparsity parameter $t=p$, the constraint $\left\lVert\bbet\right\rVert_0 \leq t$ becomes $\left\lVert\bbet\right\rVert_0 \leq p$. Since any coefficient vector $\bbet \in \mathbb{R}^p$ automatically satisfies this condition, the sparsity constraint becomes non-binding and can be removed. The RBSS subproblem then simplifies to:
	\begin{align*}
		\min_{\substack{\bbet \in \mathbb{R}^p \\ \mathcal{I}^{\prime} \subseteq \mathcal{I}}} \; \sum_{i \in \mathcal{I}^{\prime}} \left(y_i - \bx_i^T \bbet\right)^2 \quad \text{subject to} \quad |\mathcal{I}^{\prime}| \geq h.
	\end{align*}
	This is the standard formulation for the Least Trimmed Squares (LTS) estimator. \hfill $\square$

	\section*{Appendix B: Finite-Sample Breakdown Point}
		
	Theorem 1 establishes the finite-sample breakdown point (FSBP) of RBSS, introduced by \cite{thompson2022robust} and shown in Proposition 1 to be a special case of RMSS, as well as the FSBP of RMSS. 
	
	Our proofs follow closely arguments and counterexamples used in other LTS-based estimators \citep{rousseeuw1984least, rousseeuw2005robust, alfons2013sparse}, with required adaptations for the different constraints and the sparse setting of RBSS and RMSS. In our context, the number of predictors $p$ may exceed the sample size $n$ and the estimators are derived under $\ell_{0}$-sparsity constraints, e.g., to have at most $t$ nonzero components. Notably, the spatial configuration of the observations is adapted using the ``effective" dimensions of the subset of selected covariates, also referred as the ``active" set of covariates. To ease the flow and understanding of the proof, we start by introducing some useful notation and results.
	
	\vspace{.3cm}
	
	\paragraph{Notation.}
	Let $\bZ = (\bX, \by) \in \mathbb{R}^{n\times(p+1)}$ be a fixed sample of $n$ observations. As in the proof of the breakdown point of LTS, observations $\bz_i=(\bx_i^T,y_i)$ lie in a $(p+1)$-dimensional space with covariates $\bx_i \in \mathbb{R}^p$ \citep{rousseeuw1984least}. New to our setting is the definition of a restricted covariate space in $\mathbb{R}^k$ with only the coordinates corresponding to the covariates selected by a sparse estimator. More generally, for any $\bu \in \mathbb{R}^p$ and a set of indices $K \subseteq \{1,\ldots,p\}$, with $|K|=k$, denote by $\bu_K\in\mathbb{R}^{k}$ the subvector formed by the components $(u_j)_{j\in K}$. We also write $K^c=\{1,\ldots,p\}\setminus K$ and $\bu_{K^c}$ for the complementary subvector. Let $Q$ be the value of the objective function, which depends on the coefficients vector(s) and the data.
	
	\paragraph{General position for restricted subsets and assumptions.}
	
	As in LTS, we drop observations with $\bx_i=\mathbf{0}$. Since $p$ may exceed $n$ and the $\ell_0$-sparsity constraint implies only a subset of covariates is active when $t < p$, we adapt the assumption of general position to the restricted-subset case. 
	
	For fixed $ t $ and $q = \min\{2t,p\}$, with $2t < n$, let $\bX_{K}$ be the subset of $k$ covariates indexed by $K$, with $t \leq k \leq q$. We assume that the observations of any submatrix with $k$ columns of $\bX$ are in general position; i.e., no $k$ rows $\bx_{i,K} \in\mathbb{R}^k$ lie on a $(k -1)$-dimensional linear subspace through the origin.
	
	Note that these conditions do not impose any restriction on the relative size of $p$ and $n$. In our context, it is natural to assume $p>1$. Let $M_y = \max_{1 \leq i \leq n} |y_i|$, $M_{x_1} = \max_{1 \leq i \leq n} |x_{i1}|$, and $M_{x} = \max_{1 \leq j \leq p} M_{x_j}$. For a given dataset $\bZ$, these constants are fixed and finite.
	
	\paragraph{Distance equivalence.}
	For any $K\subseteq\{1,\ldots,p\}$ with $|K|=k$, let
	\[
	R_K \;=\; \{(\bx^T,y)\in\mathbb{R}^{p+1}:\ y=0,\ \bx_{K^c}=\mathbf{0}\}.
	\]
	Since the mapping from $R_K$ to $\mathbb{R}^k$ that retains only the coordinates in $K$ preserves Euclidean distances, distances computed within $R_K$ are identical to those computed in $\mathbb{R}^k$.
	
	\paragraph{Proof of Theorem 1-(I): FSBP of RBSS.}
	
	Suppose $\hat{\bbet} = \hat{\bbet}(\bZ)$ minimizes \eqref{eq:RBSS}, i.e., $\hat{\bbet}$  is the RBSS estimator of coefficients for $\bZ$. Let $\hat{\mathcal{I}}$ be the indices corresponding to the $h$ smallest residuals. Define $M = \max_{1 \leq i \leq n} |y_i - \bx_i^T\hat{\bbet}|$. Let $\left[ \frac{n+k+1}{2} \right] \le h \le n$, we show that $\varepsilon^*(\hat{\bbet}; \bZ) = \tfrac{n-h + 1}{n}$ by establishing lower and upper bounds that coincide.
	
	\vspace{.5cm}
	
	\noindent \textbf{Lower bound}: $\varepsilon^*(\hat{\bbet}; \bZ) \geq \tfrac{n-h + 1}{n}$. This part of the proof mirrors the LTS proof in \citet{rousseeuw2005robust}, except for new, tailored arguments addressing the selection-dependent subspaces induced by the $\ell_0$-sparsity constraint. Define $S = \operatorname{supp}(\hat{\bbet}) = \{\, j : \hat{\beta}_j \neq 0 \,\}$, which also identifies the active set of predictors. Let $H = \{(\bu, \bu^t \hat{\bbet}) \in \mathbb{R}^{(p+1)}\}$ be the nonvertical hyperplane defined by $\hat{\bbet}$. 
	
	Let $\tilde{\bZ}$ be a contaminated sample resulting from replacing $m \leq n-h $ observations by arbitrary observations. Then,  the contaminated sample $\tilde{\bZ}$ contains $n-m \geq h$ of the original observations from $\bZ$, which we call the ``good" observations. For simplicity, let $\tilde{\bbet} = \hat{\bbet}(\tilde{\bZ})$ be the minimizer of RBSS problem \eqref{eq:RBSS} for the contaminated sample $\tilde{\bZ}$. Let $\tilde{\mathcal{I}}$ be the indices corresponding to the $h$ smallest residuals of $\tilde{\bZ}$ with respect to $\tilde{\bbet}$. Note that $\tilde{\bZ}$ and $\bZ$ share more than $h$ observations, thus the sum of the smallest $h$ squared residuals of points in $\tilde{\bZ}$ with respect to $\hat{\bbet}$ will contain such points with the residuals as before. Thus,
	
	\begin{equation}
		\sum_{i \in \tilde{\mathcal{I}}}(\tilde{y}_i - \tilde{\bx}_i^T \tilde{\bbet})^2 \leq \sum_{i \in \tilde{\mathcal{I}'}}(\tilde{y}_i - \tilde{\bx}_i^T \hat{\bbet})^2 \leq hM^2,
		\tag{A.1}
	\end{equation}

	Define $\tilde{S}=\operatorname{supp}(\tilde{\bbet})$ and let $\tilde{H} = \{(\bu, \bu^t \tilde{\bbet}) \in \mathbb{R}^{(p+1)}\}$ be the corresponding hyperplane for $\tilde{\bbet}$. Without loss of generality, assume that $\tilde{\bbet} \neq \hat{\bbet}$, so that $H \neq \tilde{H}$.
	
	Let $K \;=\; S\cup \tilde{S}, \; |K| = k$ be the indices of all selected variables by both estimators, where $ t \leq k \leq \min\{2t, p\}$ given the $\ell_{0}$-sparsity constraint and potential overlap between the estimators. Consider the restricted set of covariate $\bX_K \in \mathbb{R}^{n \times k}$ with only the covariates selected by the sparse estimators, indexed by $K$. Given our adapted assumptions, the observations in this submatrix are in general position, so there exist a positive constant $\rho$ such that
	\[
	\rho=\tfrac12 \inf 
	\begin{aligned}[t]
		\{&\tau>0:\ \text{there exists a }(k-1)\text{-dimensional subspace } V \text{ of }\mathbb{R}^k \text{ through the origin}\\
		& \text{such that } V^\tau \text{ covers at least } k \text{ of the } \bx_{i,K}\},
	\end{aligned}
	\]
	
	\noindent where $V^\tau$ is the set of all points $\bu$ with distance to V not larger than $\tau$. 
	
	Regardless of the selection of covariates, $\dim(H \cap \tilde{H})$ is $p-1$. However, all the covariates indexed by $K^c$ are not selected by neither of the estimators and those coordinates can be dropped when measuring distances in the restricted subspaces. More formally, let $R_K = (y=0)\cap\{x_{K^c}=\mathbf{0}\}$. Projecting $H \cap \tilde{H}$ onto $(y=0)$ as in the proof of LTS \citep{rousseeuw2005robust} is equivalent to project it onto $R_K$, 
	
	$$P=\{\bu \in\mathbb{R}^p:\ \bu^T\Delta=\mathbf{0}\} =\{\bu \in\mathbb{R}^p:\ \bu_K^T\Delta_K=\mathbf{0}\},$$
	
	\noindent where $\Delta = \tilde{\bbet} - \hat{\bbet}$ and $\Delta_K$ be the $k$-dimensional vector containing only the components of $\Delta$ with indices in $K$. We note that measuring distances in $P$ is equivalent to measure them in the $(k-1)$-dimensional hyperplane in the restricted space
	
	\[
	P_K \;=\; \{\bv \in\mathbb{R}^{k}:\ \bv^T\Delta_K=\mathbf{0}\}.
	\]
	
	Since we assumed that for any $k$ subset of columns of $\bX$, points are in general position, then at most $k-1$ of the good points $\bx_{i,K}$ lie at a distance less than $\rho$ from $P_K$ and we define $A$ as the set of the remaining good observations. 
	
	From here, the proof follows identically to the proof for LTS, measuring distances in the restricted covariate subspace and replacing $p$ by $k$, which characterizes the ``effective" set size of the restricted subspace. In particular, for any point good point $(\bx^T_a, y_a)$ in $A$,
	
	$$
	|\tilde{r}_a-r_a|
	=\big|\boldsymbol{x}_a^T(\tilde{\bbet}-\hat{\bbet})\big|
	=\big|\boldsymbol{x}_{a,K}^T\Delta_K\big|
	> \rho(\|\tilde{\bbet}-\hat{\bbet}\|_2 - 2\|\hat{\bbet}\|_2 ),
	$$
	
	\noindent where $\tilde{r}_a$ and $r_a$ are the residuals of the $(\bx^T_a, y_a)$ to $\tilde{\bbet}$ and $\hat{\bbet}$, respectively. 
	
	Assume that 
	
	$$\bigl\|\hat{\bbet} - \tilde{\bbet}\bigr\|_2 \geq 2\bigl\|\hat{\bbet}\bigr\|_2 + M (1 + \sqrt{h})/ \rho).$$
	
	Then, for all points in $A$,
	
	$$
	|\tilde{r}_a| \geq |\tilde{r}_a-r_a| - |r_a| >
	\rho(\|\tilde{\bbet}-\hat{\bbet}\|_2 - 2\|\hat{\bbet}\|_2 ) - M  \geq M (1 + \sqrt{h}) - M = \sqrt{h} M
	$$
	
	Since any set of $h$ points of $\tilde{\bZ}$ must contain at least one of the points of $A$, then
	
	$$
	\sum_{i \in \tilde{I}}(\tilde{y}_i - \tilde{\bx}_i^T \tilde{\bbet})^2 \geq (\tilde{r}_a)^2 > hM^2,
	$$
	
	\noindent which contradicts (A.1). Thus,
	
	$$\bigl\|\hat{\bbet} - \tilde{\bbet}\bigr\|_2 < 2\bigl\|\hat{\bbet}\bigr\|_2 + M (1 + \sqrt{h})/ \rho) < \inf$$
	
	\noindent for all contaminated samples $\tilde{\bZ}$. \hfill $\square$
	
	\vspace{0.5cm}
	
	\noindent \textbf{Upper bound}: $\varepsilon^*(\hat{\bbet}; \bZ) \leq \tfrac{n-h+1}{n}$. In this part, we show that we can break down RBSS estimator if $n-h+1$ observations of $\bZ$ are contaminated. Slightly adapting arguments from the proof of SparseLTS' breakdown point (\cite{alfons2013sparse}), we construct an adversarial contaminated sample for which RBSS would not stay bounded. 
	
	Let $\tilde{\bZ}_{\gamma,\tau}$ be a contaminated sample obtained by replacing $m = n-h+1$ observations of $\bZ$ with the common contaminated point $\bz_c = (\bx_c^T, y_c)^T$, defined by arbitrary positive constants $\gamma$ and  $\tau$:
	\[
	\bx_c = (\tau, 0, \ldots, 0)^T \in \mathbb{R}^p \; \text{and} \; y_c = \gamma \tau.
	\]
	
	\noindent Note that $\tilde{\bZ}_{\gamma,\tau}$ and $\bZ$ share exactly $h-1$ observations, which we consider ``good" points, and $n-h+1$ identical contaminated points. 
	
	Assume that the RBSS estimator based on any such contaminated sample $\tilde{\bZ}_{\gamma,\tau}$ does not break down, i.e., there exists a constant $M$ such that
	
	\begin{equation}
		\sup_{\tau,\gamma} \|\hat{\bbet}(\tilde{\bZ}_{\gamma,\tau})\|_2 \leq M. 
		\tag{B.1}
	\end{equation}

	We next show that this assumption results in a contradiction. In particular, consider $\gamma = M+2$ and $\tau > 0$ such that
	\[
	\tau^2 \geq \max(h-m,0)(M_y + \gamma M_{x_1})^2 + 1.
	\]
	
	\noindent For this value of $\gamma$, let $\bbet_\gamma = (\gamma, 0, \ldots, 0)^\prime \in \mathbb{R}^p$ be a sparse estimate candidate. Note that $\left\lVert \bbet_\gamma \right\rVert_0 \leq t$ so this is a feasible candidate in the minimization of the objective function. Since the residuals of the $m=n-h+1$ contaminated points are all zero, these points will be part of the set of $h$ lowest squared residuals used to minimize the objective function $Q$. If $h > m$, the objective function will also sum the squared of $h-m = 2h - n -1$ ``good" points in $\tilde{\bZ}_{\gamma,\tau}$ shared with the original sample. Thus, the resulting value of objective function for $\bbet_\gamma$ becomes
	\[
	Q(\bbet_\gamma, \tilde{\bZ}_{\gamma,\tau}) =
	\begin{cases}
		\displaystyle \sum_{i= n-h+2}^{h} \big((y - \bX^T\bbet_\gamma)^2\big)_{i:n} 
		, & \text{if } h > m, \\[2ex]
		0, & \text{else},
	\end{cases}
	\]
	
	\noindent where $0 \leq \big(r^2\big)_{(n-h+2):n} \leq \ldots \leq \big(r^2\big)_{n:n}$ are the ordered squared residuals for all points in $\tilde{\bZ}_{\gamma,\tau}$ relative to $\bbet_\gamma$. Hence,
	\[
	Q(\bbet_\gamma, \tilde{\bZ}_{\gamma,\tau}) \leq \max(h-m,0)(M_y + \gamma M_{x_1})^2  
	\leq \tau^2 - 1 < \tau^2. 
	\]
	
	\noindent Given that the value of the objective function evaluated at the minimizer is smaller than that at any other candidate estimate, then
	\begin{equation}
		Q(\hat{\bbet}, \tilde{\bZ}_{\gamma,\tau}) \leq Q(\bbet_\gamma, \tilde{\bZ}_{\gamma,\tau}) < \tau^2.
		\tag{B.2}
	\end{equation}
	
	Now consider any feasible candidate $\bbet$ such that $\left\lVert \bbet \right\rVert_0 \leq t$ and $\left\lVert \bbet \right\rVert_2 \leq \gamma-1$. Since at least one contaminated point must be in the set of the smallest $h$ residuals, then its objective function sums the squared residual of such point, i.e.,
	\[
	Q(\bbet, \tilde{\bZ}_{\gamma,\tau}) \geq (\gamma \tau - \tau \beta_1)^2 \geq \max(\tau^2, (\tau \gamma)^2) \geq \tau^2,
	\]
	
	\noindent since $\beta_1 \leq \|\bbet\|_2 \leq \gamma - 1$. Then, from (B.2) it follows that
	\[
	\|\hat{\bbet}(\tilde{\bZ}_{\gamma,\tau})\|_2 > \gamma - 1 = M+1,
	\]
	\noindent which contradicts (B.1). Thus, RBSS breaks down under this level of contamination. \hfill $\square$
	
	\paragraph{Proof of Theorem 1-(II): FSBP of RMSS.}
	
	Let $\hat{\calsB} = \hat{\calsB}(\bZ) = \{\hat{\bbet}^1(\bZ), \dots, \hat{\bbet}^G(\bZ)\}$ be the RMSS solution of \eqref{eq:RMSS} for the clean data $\bZ$. Since $\hat{\calsB}$ is a set of $G$ vectors of model coefficients, we can think that it breaks down if at least one of the vector of coefficients $\hat{\bbet}^g$ breaks down. Thus, our objective is to show that $\varepsilon^*(\hat{\bbet}^g; \bZ) = \frac{n - h + 1}{n}$ for all $g: 1 \leq g \leq G$. As in part (I), the proof establishes equal lower and upper bounds for the breakdown point of RMSS. Because each $\hat{\bbet}^g$ defines a submodel, we use the terms coefficient vector and submodel interchangeably. Define $M_g = \max_{1 \leq i \leq n} |y_i - \bx_i^T \hat{\bbet}^g|$, and $M = \max_{1 \leq g \leq G} M_g$, which for a given data $\bZ$ and RMSS solution $\hat{\calsB}$ are fixed and finite. 
	
	\vspace{.5cm}
	
	\noindent \textbf{Lower bound}: For all $g:\; 1 \leq g \leq G, \; \varepsilon^*(\hat{\bbet}^g; \bZ) \geq \tfrac{n-h + 1}{n}$. We first show that all models $\hat{\bbet}^g \in \hat{\calsB}, \; 1 \leq g \leq G$, remain bounded if $m \leq n-h$ observations are arbitrarily contaminated. Assuming that that is not true, will yield a contradiction. The arguments are very similar to those used in part (I) but adapted to the simultaneous minimization over $G$ vectors of  coefficients.
	
	Let $\tilde{\bZ}$ be a contaminated sample obtained by replacing $m \le n-h$ observations. This ensures that $\tilde{\bZ}$ contains at least $n-m \ge h$ ``good" observations from $\bZ$. For simplicity, let $\tilde{\calsB} = \hat{\calsB}(\tilde{\bZ})= \{\hat{\bbet}^1(\tilde{\bZ}), \dots, \hat{\bbet}^G(\tilde{\bZ})\}$ be the RMSS solution based on the contaminated data and $\tilde{\mathcal{I}}^g$ be the indexes of the $h$ smallest residuals for each submodel.

	Assume that one of the $G$ coefficient vectors of RMSS breaks down under this contamination. Without loss of generality, let $\tilde{\bbet}$ be such model, so there exist at least one model $\hat{\bbet}$ in $\hat{\calsB}$ such that
	
	\[
	\lVert \tilde{\bbet} - \hat{\bbet} \rVert \geq 2 \lVert  \hat{\bbet} \rVert 
	+ M \Big( 1 + \sqrt{hG} \Big)/ \rho.
	\]
	
	\noindent for $\rho > 0$ defined in part (I) with the adapted assumption of points in general position for any submatrix of $k$ covariates of $\bX$. To simplify the notation, we omit the superscripts that identify the specific submodels and the reference to the data employed in the estimation. Note that all coefficient vectors of $\hat{\calsB}$ are feasible solutions of \eqref{eq:RMSS}, so for the contaminated sample, the value of its objective function based on $\tilde{\bZ}$ is above that of the minimum. Moreover, given that more than $h$ points of $\bZ$ remained unchanged in $\tilde{\bZ}$, the sum of the smallest residuals of the new sample $\tilde{\bZ}$ with respect to each old submodel $\hat{\bbet}^g$, indexed by $\tilde{\mathcal{I}'^g}$, is less than or equal to $h(M_{g})^2$, thus
	
	\[
	\sum_{g = 1}^G \sum_{i \in \tilde{\mathcal{I}^g}} \big(\tilde{y}_i - \tilde{\bx}_i^T \tilde{\bbet}^g \big)^2
	\;\;\leq\;\;
	\sum_{g = 1}^G \sum_{i \in \tilde{\mathcal{I}'^g}} \big(\tilde{y}_i - \tilde{\bx}_i^T \hat{\bbet}^g \big)^2
	\;\;\leq\;\;
	\sum_{g = 1}^G h M_g^2
	\;=\; G h M^2.
	\]

	To reach a contradiction, we now focus on the hyperplanes $H = \{(\bx, \bx^t \hat{\bbet})\}$, $\tilde{H} = \{(\bx, \bx^t \tilde{\bbet})\}$, their intersection and their projection onto the subspace $R_K = (y=0)\cap\{x_{K^c}=0\}$, where $K^c$ is the complement of the subset $K$ (i.e., indices of non-selected variables). 
	
	Based on our extended assumption of points in general position and following the same arguments as in \cite{rousseeuw2005robust} and part (I), we can show that for any ``good'' point $(\bx^T_a,y_a)$ in $A$,
	\[
	|\tilde{r}_a - r_a | 
	\;>\; 
	\rho \big( \lVert \tilde{\bbet} - \hat{\bbet} \rVert - 2 \lVert \hat{\bbet} \rVert \big).
	\]
	
	\noindent where $\tilde{r_a}$ and $r_a$ represent the residuals of $(\bx^T_a,y_a)$ to $\tilde{\bbet}$ and $\hat{\bbet}$, respectively. Then, 
	\[
	| \tilde{r}_a | \;\geq\; | \tilde{r}_a  - r_a | - | r_a |
	\;>\; (1+\sqrt{hG})M - M = \sqrt{Gh}M.
	\]
	
	Since any set of $h$ points of $\tilde{\bZ}$, including $\tilde{\mathcal{I}}$, must contain at least one of these $(y_a, \bx_a)$, then 
	
	\[
	\sum_{g = 1}^G  \sum_{i \in \tilde{\mathcal{I}}^g} (\tilde{y}_i - \tilde{\bx}_i^T \tilde{\bbet}^g)^2
	\;\;\geq\;\;
	\sum_{i \in \tilde{\mathcal{I}}} (\tilde{y}_i - \tilde{\bx}_i^T \tilde{\bbet})^2
	\;\;\geq\;\; (\tilde{r}_a)^2 \;\;>\;\; GhM^2,
	\]
	
	\noindent which is a contradiction. Then, 
	
	\[
	\lVert \tilde{\bbet} - \hat{\bbet} \rVert < 2 \lVert  \hat{\bbet} \rVert 
	+ M \Big( 1 + \sqrt{hG} \Big)/ \rho < \inf,
	\]
	
	\noindent for all contaminated samples $\tilde{\bZ}$. \hfill $\square$
	
	\vspace{0.5cm}
	
	\noindent \textbf{Upper bound}: For all $g:\; 1 \leq g \leq G, \; \varepsilon^*(\hat{\bbet}^g; \bZ) \leq \tfrac{n-h + 1}{n}$. As before, we can construct an adversarial dataset with $m = n-h+1$ contaminated points that forces at least one model in the RMSS solution to be unbounded. Both the example and the arguments are very similar to those used in part (I).
	
	Let's first consider the case that $G < p$, and let $\tilde{\bZ}_{\gamma,\tau}$ be a sample obtained by replacing $m=n-h+1$ observations with an identical point $\bz_c = (\bx_c^T, y_c)^T$, where $\bx_c = (\tau, \tau, \ldots, \tau)^T$ and $y_c = \gamma \tau$ for any positive constants $\gamma$ and $\tau$. 
	
	Assume that the RMSS estimator based on any such contaminated sample $\tilde{\bZ}_{\gamma,\tau}$ does not break down, i.e., there exists a constant $M$ such that
	
	\begin{equation}
		\sup_{\tau,\gamma} \|\hat{\bbet}^g(\tilde{\bZ}_{\gamma,\tau})\|_2 \leq M, \; 
		\forall\, g \;:\; 1 \leq g \leq G
		\tag{C.1}
	\end{equation}

	We next show that this assumption results in a contradiction. In particular, consider $\gamma = M+2$ and $\tau > 0$ such that
	
	\[
	\tau^2 \geq \max(h-m,0)(M_y + \gamma M_{x})^2 + 1.
	\]
	
	\noindent For this value of $\gamma$, let $\calsB_{\gamma} = \{\bbet^1_{\gamma}, \dots, \bbet^G_{\gamma}\}$, a candidate solution for the RMSS problem \eqref{eq:RMSS} based on the contaminated data $\tilde{\bZ}_{\gamma,\tau}$, where the $i$-th estimate, $1 \leq j \leq t$, of each submodel's coefficient vector $\bbet_\gamma^g$ is defined as $\bbet^g_{\gamma, j} = \gamma$ if $j = g, \; 1 \leq g \leq G$ and $0$ otherwise.
	
	\noindent Note that this candidate set $\calsB_{\gamma}$ is feasible under the RMSS constraints. The sparsity constraint $||\bbet^g_\gamma||_0 \le t$ is met for $t \ge 1$. The diversity constraint is also met, as for predictor $j=1$, $||\bbet_{\gamma, j,\cdot}||_0 = 1 \le u$ (for $u \ge 1$), and for all $1 \leq j \leq G$.

	As before, and by construction, the residuals of the $m=n-h+1$ contaminated points for any submodel defined by $\calsB_{\gamma}$ are all zero. Thus, these contamination points will be part of all $G$ sets of $h$ lowest squared residuals used to minimize the objective function $Q$. If $h > m$, the objective function will also sum the squared of $h-m$ ``good" points in $\tilde{\bZ}_{\gamma,\tau}$ shared with $\bZ$. Thus, the resulting value of objective function for $\calsB_{\gamma}$ has a similar upper bound as before, although in this case we are analyzing all $G$ models simultaneously, 
	
	\[
	Q(\calsB_{\gamma}, \tilde{\bZ}_{\gamma,\tau}) \leq G\max(h-m,0)(M_y + \gamma M_{x})^2  
	\leq G(\tau^2 - 1) < G\tau^2. 
	\]
	
	\noindent Thus, 
	
	\begin{equation}
		Q(\hat{\calsB}, \tilde{\bZ}_{\gamma,\tau}) \leq Q(\calsB_{\gamma}, \tilde{\bZ}_{\gamma,\tau}) < \tau^2.
	\end{equation}
	
	Consider any feasible set of coefficient vectors $\calsB = (\bbet^1, \ldots,\bbet^G)$ such that $\left\lVert \bbet^g \right\rVert_2 \leq \gamma-1, \; \forall\, g \;:\; 1 \leq g \leq G$, and assume that other constrains are met. Since at least one outlier must be in the set of the smallest $h$ residuals of all solutions, then for any such $\calsB$ its objective function sums the squared of the residual of at least one contaminated point to each coefficient vector $\bbet^g$, for all $1\leq g \leq G$. So,
	
	\[
	Q(\calsB, \tilde{\bZ}_{\gamma,\tau}) \geq \sum_{g=1}^G \left(\gamma \tau - \tau \sum_{j = 1}^p\beta^g_j\right)^2 \;\;\geq\;\; \sum_{g=1}^G\tau^2 \left( \gamma - \lVert \bbet^g \rVert_2 \right)^2 \geq G \tau^2.
	\]

	\noindent Then,
	
	\[
	\|\hat{\bbet}^g(\tilde{\bZ}_{\gamma,\tau})\|_2 > \gamma - 1 = M+1,
	\]
	\noindent which contradicts the assumption (C.1) and shows that all model estimators break down under this level of contamination.  
	
	In our manuscript, we focus on the case where $G < p$ and $p >>n$. However, if $G \geq p$, we can adapt the contaminated sample accordingly. In that case, the models need to share variables, i.e., the constraint $u>1$. In such case, more than one coordinate of each $\bbet_\gamma$ can equal to $\gamma$ and we can adapt the value of $y_c$ accordingly so that the residuals continue to be zero.
	
	\hfill $\square$
	
	\section*{Appendix C: Combinatorial Complexity of RMSS}
	
	A brute-force approach to solving the RMSS optimization problem would require evaluating every possible combination of predictor subsets and observation subsets. In this section, we demonstrate that this is computationally infeasible, even for small problems and for the simplest case where predictors are not shared between models ($u=1$).
	
	Let $p_g$ be the number of predictors in model $g$, $1 \leq g \leq G$, with the constraint that $p_g \leq t$. The total number of unique ways to partition the predictors and select observations for an RMSS ensemble with $u=1$ is given by the product of two terms: the number of ways to choose the observation subsets and the number of ways to partition the predictors.
	Let $q = \sum_{g=1}^G p_g$, and let $h_j(p_1, \dots, p_G)$ be the number of models with exactly $j$ predictors. The total number of combinations is:
	\begin{equation} \label{eq:combinatorics}
		\underbrace{\left[\sum_{i=h}^n \binom{n}{i}\right]^G}_{\text{Observation Subset Choices}} \times \underbrace{\sum_{p_1 \leq \dots \leq p_G \leq t} \binom{p}{q} \left[\frac{q!}{p_1! \dots p_G!} \prod_{j=1}^{t} \frac{1}{h_j(p_1, \dots, p_G)!}\right]}_{\text{Predictor Partition Choices}}.
	\end{equation}
	The first term counts the number of ways to independently choose at least $h$ observations for each of the $G$ models. The second term, a sum over all possible model sizes, counts the number of ways to partition $p$ predictors into $G$ disjoint subsets of sizes $p_g$, $1 \leq g \leq G$.
	
	The combinatorial explosion demonstrated by this formula makes a brute-force evaluation intractable. To illustrate, consider a simple, low-dimensional case with $n=10$ samples, $p=10$ predictors, and $G=2$ models. If we fix the tuning parameters to $t=5$, $h=5$, and $u=1$, formula \eqref{eq:combinatorics} reveals there are over 9 trillion unique combinations for the two models. This staggering number, for a problem of trivial size, underscores the necessity of the efficient computational algorithm proposed in the main article. The challenge is magnified exponentially when the tuning parameters $t$, $u$, and $h$ must themselves be selected via cross-validation over a grid of candidate values.

	\section*{Appendix D: Robust Stepwise Algorithm}
	
	In Section \ref{sec:stepwise_lemmas}, we begin by outlining the two main lemmas from \cite{khan2007building}, which establish that the stepwise search algorithm can be expressed solely in terms of correlations. For completeness and to improve the clarity of the exposition, in this Section we provide full proofs of Lemmas 1 and 2 from \citet{khan2007building}. We then provide the proof for Proposition 2 of the main article. This proposition presents elegant explicit formulas that rely exclusively on the correlation matrix of the predictors, as well as the correlation between the predictors and the response variable. The formulas, consistent with our algorithm and code, enable us to identify the optimal predictor at each step for a given model, along with the robust residual sum-of-squares (rRSS) required for the robust partial F-rules.
	
	\subsection*{The  Lemmas} \label{sec:stepwise_lemmas}

	In each step of the classical (non-robust) single-model stepwise regression algorithm, one must identify the predictor variable that reduces the residual sum of squares (RSS) by the largest amount when combined with variables already in the model, and test whether this reduction is statistically significantly with respect to some threshold $ \gamma \in (0, 1) $ via a partial $ F $-test \citep[see e.g.][]{pope1972use}. Let $ \bX \in \mathbb{R}^{n \times p} $ be the matrix for $ n $ samples of $ p $ covariates and let $ \by \in \mathbb{R}^n $ be the vector of response variables. We assume that the variables have been centered to zero and scaled to one. In a robust setting, the variables can be centered using the median and scaled using the MAD (median absolute deviation from the median). 
	
	Denote $\mathbf{\hat{r}_{y}} = (\hat{r}_{1y}, \dots, \hat{r}_{py})^T \in \mathbb{R}^p$ the vector of sample correlations between the response and predictor variables, and $ \mathbf{\hat{\Sigma}} \in \mathbb{R}^{p \times p} $ the sample correlation matrix of $ \bX$. For notational convenience we denote the set $ \mathcal{J} = \{1, \dots, p\} $ and $ \bX_j $ the $ j $-th column of the design matrix. The following lemmas  proved by induction in \cite{khan2007building}  form the basis for the robustification of stepwise regression. Since Proposition 2 builds on Lemmas 1 and 2 from \citet{khan2007building}, we include their complete proofs here for clarity.
	
	\begin{lemma} \label{lemma:step1}
		Only the original correlations $ \mathbf{\hat{r}_{y}} $ and $ \mathbf{\hat{\Sigma}} $ are needed to generate the sequence of variables from a stepwise regression search algorithm.
	\end{lemma}
	\begin{lemma} \label{lemma:step2}
		The partial $ F $-test at each step of a stepwise regression search algorithm for nested model comparison can be written as a function of $ \mathbf{\hat{r}_{y}} $ and $ \mathbf{\hat{\Sigma}} $ only.
	\end{lemma}
	The stepwise algorithm can thus be robustified by replacing the classical estimates of $ \mathbf{\hat{r}_{y}} $ and $ \mathbf{\hat{\Sigma}} $ by their robust counterparts, which we obtain via the DDC (detecting deviating cells) method of \cite{rousseeuw2018detecting}. The latter method can be scaled efficiently to high-dimensional settings \citep{raymaekers2021fast}.
	
	\paragraph{Proof of Lemma 1.}
	We first derive the first three steps of the forward stepwise algorithm to add  variables to the null model using partial correlations between variables,  and generalize the computations for any step of the algorithm.
	\begin{itemize}
		\item Variable $ j_1 \in \mathcal{J} $ maximizes the correlation $ r_{jy} $,
		\begin{align*}
			j_1 = \argmax_{j \in \mathcal{J}}  \left| \hat{r}_{jy} \right|,
		\end{align*}
		so $ j_1 $ enters the model first.
		\item Denote the residual vector $\bZ_{j.j_1} = X_j - \beta_{j j_1} X_{j_1}$ where $\beta_{j j_1} = \mathbf{\hat{\Sigma}}_{j j_1}$, $j \in \mathcal{J} \setminus \{j_1\}$. Since $X_{j_1}$ is standardized to have mean 0 and standard deviation 1, each residual vector $\bZ_{j.j_1}$ maintains mean 0. Variable $j_2$ maximizes the (scaled) partial correlation,
		\begin{align*}
			j_2 &= \argmax_{j \in \mathcal{J} \setminus\{j_1\}} \left| \frac{\bZ_{j.j_1}^T \left(\by - \beta_{j j_1} X_{j_1} \right)/n}{\sqrt{\bZ_{j.j_1}^T\bZ_{j.j_1}/n} \cdot \text{SD}\left(\by - \beta_{j j_1} X_{j_1} \right)} \right| \\
			&= \argmax_{j \in \mathcal{J} \setminus\{j_1\}} \left| \frac{\bZ_{j.j_1}^T \by/n}{\sqrt{\bZ_{j.j_1}^T\bZ_{j.j_1}/n}} \right|,
		\end{align*}
		where the denominator term $\text{SD}\left(\by - \beta_{j j_1} X_{j_1} \right)$ is removed in the second line because it does not depend on the covariate index $j$ being optimized over. It represents the standard deviation of the response residual after removing the effect of $X_{j_1}$, which is constant across all candidate covariates. Thus $j_2$ enters the model second.
		\item Denote the residual vector $ \bZ_{j.j_1j_2} = \bZ_{j.j_1} - \beta_{jj_2.j_1} \bZ_{j_2.j_1}, $ where
		\begin{align*}
			\beta_{jj_2.j_1}= \frac{\bZ_{j_2.j_1}^T\bZ_{j.j_1}}{\bZ_{j_2.j_1}^T\bZ_{j_2.j_1}}, \quad j \in \mathcal{J} \setminus\{j_1, j_2\}.
		\end{align*}
		Variable $ j_3 $ maximizes the (scaled) partial correlation,
		\begin{align*}
			j_3 = \argmax_{j \in \mathcal{J}\setminus\{j_1, j_2\}} \left| \frac{\bZ_{j.j_1j_2}^T \by/n}{\sqrt{\bZ_{j.j_1j_2}^T\bZ_{j.j_1j_2}/n}} \right|,
		\end{align*}
		so $ j_3 $ enters the model third.
	\end{itemize}
	For steps $ k \geq 2 $, we can generalize the forward stepwise in the following way. Denote the residual vector $ \bZ_{j.j_1\dots j_{k-1}} = \bZ_{j.j_1\dots j_{k-2}} - \beta_{jj_{k-1}.j_1\dots j_{k-2}} \bZ_{j_{k-1}.j_1 \dots j_{k-2}}, $ where
	\begin{align*}
		\beta_{jj_{k-1}.j_1\dots j_{k-2}}= \frac{\bZ_{j_{k-1}.j_1\dots j_{k-2}}^T\bZ_{j.j_1\dots j_{k-2}}}{\bZ_{j_{k-1}.j_1\dots j_{k-2}}^T\bZ_{j_{k-1}.j_1\dots j_{k-2}}}, \quad j \in \mathcal{J} \setminus\{j_1, \dots, j_{k-1}\}.
	\end{align*}
	At step $ k \geq 2$, variable $ j_k $ maximizes the (scaled) partial correlation,
	\begin{align*}
		j_k = \argmax_{j \in \mathcal{J}\setminus\{j_1, \dots, j_{k-1}\}} \left| \frac{\bZ_{j.j_1\dots j_{k-1}}^T \by/n}{\sqrt{\bZ_{j.j_1\dots j_{k-1}}^T\bZ_{j.j_1\dots j_{k-1}}/n}} \right|,
	\end{align*}
	so $ j_k $ enters the model third. Thus we only need to show that for any $ k \geq 2 $:
	\begin{enumerate}
		\item $ \bZ_{j_{k}.j_1\dots j_{k-1}}^T\bZ_{j_{k}.j_1\dots j_{k-1}} $, and
		\item $ \bZ_{j.j_1\dots j_{k-1}}^T \by $
	\end{enumerate}
	can be written as a function of $ \mathbf{\hat{\Sigma}} $ and $\mathbf{\hat{r}}_{\by}$ only.
	
	First, note that
	\begin{align*}
		\bZ_{j_2.j_1}^T \bZ_{j.j_1}= \left[\bX_{j_2} - \mathbf{\hat{\Sigma}}_{j_1 j_2} \bX_{j_1} \right]^T\left[\bX_{j} - \mathbf{\hat{\Sigma}}_{j_1 j} \bX_{j_1} \right] = \mathbf{\hat{\Sigma}}_{j j_2} - \mathbf{\hat{\Sigma}}_{j j_1}  \mathbf{\hat{\Sigma}}_{j_1 j_2}.
	\end{align*}
	For $ k \geq 3 $:
	\begin{align*}
		\bZ_{j_k.j_1\dots j_{k-1}}^T \bZ_{j.j_1\dots j_{k-1}}  &= \left[\bZ_{j_k.j_1\dots j_{k-2}} - \beta_{j_k j_{k-1}.j_1\dots j_{k-2}} \bZ_{j_{k-1}.j_1 \dots j_{k-2}} \right]^T \\
		&\quad\quad \left[ \bZ_{j.j_1\dots j_{k-2}} - \beta_{j j_{k-1}.j_1\dots j_{k-2}} \bZ_{j_{k-1}.j_1 \dots j_{k-2}}   \right] \\
		&= \bZ_{j_k.j_1\dots j_{k-2}}^T \bZ_{j.j_1\dots j_{k-2}} \\
		&\quad\quad - \beta_{j j_{k-1}.j_1\dots j_{k-2}} \bZ_{j_{k}.j_1\dots j_{k-2}}^T \bZ_{j_{k-1}.j_1\dots j_{k-2}} \\
		& \quad \quad - \beta_{j_k j_{k-1}.j_1\dots j_{k-2}} \bZ_{j_{k-1}.j_1\dots j_{k-2}}^T \bZ_{j.j_1\dots j_{k-2}} \\
		&\quad\quad + \beta_{j j_{k-1}.j_1\dots j_{k-2}} \beta_{j_k j_{k-1}.j_1\dots j_{k-2}}  \bZ_{j_{k-1}.j_1\dots j_{k-2}}^T \bZ_{j_{k-1}.j_1\dots j_{k-2}},
	\end{align*}
	where 
	\begin{align*}
		\beta_{jj_{k-1}.j_1\dots j_{k-2}}= \frac{\bZ_{j_{k-1}.j_1\dots j_{k-2}}^T\bZ_{j.j_1\dots j_{k-2}}}{\bZ_{j_{k-1}.j_1\dots j_{k-2}}^T\bZ_{j_{k-1}.j_1\dots j_{k-2}}}, \quad j \in \mathcal{J}\setminus\{j_1, \dots, j_{k-1}\}.
	\end{align*}
	If the inner products $ \bZ_{j_{a}.j_1\dots j_{k-2}}^T\bZ_{j_b.j_1\dots j_{k-2}} $ can be written as a function of $ \mathbf{\hat{\Sigma}} $ and $\mathbf{\hat{r}}_{\by}$ for $j_a, j_b \in \mathcal{J}\setminus\{j_1, \dots, j_{k-1}\}$, then the inner products $ \bZ_{j_a.j_1\dots j_{k-1}}^T \bZ_{j_b.j_1\dots j_{k-1}} $ can be written as a function of $ \mathbf{\hat{\Sigma}} $ and $\mathbf{\hat{r}}_{\by}$. Since 	$ \bZ_{j_a.j_1}^T \bZ_{j_b.j_1} = \mathbf{\hat{\Sigma}}_{j_a j_b} - \mathbf{\hat{\Sigma}}_{j_1 j_a}  \mathbf{\hat{\Sigma}}_{j_1 j_b} $, this proves the first part of Lemma 1.
	
	Second, note that
	\begin{align*}
		\frac{1}{n}\bX_{j_1}^T \by &= \hat{r}_{y j_1}, \\
		\frac{1}{n}\bZ_{j.j_1}^T \by &= \left[\bX_{j} - \mathbf{\hat{\Sigma}}_{j j_1} \bX_{j_1} \right]^T \by = \hat{r}_{y j} - \mathbf{\hat{\Sigma}}_{j j_1} \hat{r}_{y j_1}, \quad j \in \mathcal{J}\setminus\{j_1, \dots, j_{k-1}\}.
	\end{align*}
	For $ k \geq 2 $,
	\begin{align*}
		\bZ_{j.j_1 \dots j_k}^T \by = \bZ_{j.j_1\dots j_{k-1}}^T \by - \beta_{j j_k.j_1 \dots j_{k-1}} \bZ_{j_k.j_1 \dots j_{k-1}}^T \by, \quad j \in \mathcal{J}\setminus\{j_1, \dots, j_{k-1}\}.
	\end{align*}
	Since it was already proved that $ \beta_{j j_k.j_1 \dots j_{k-1}} $ can be written as a function of $ \mathbf{\hat{\Sigma}} $ and $\mathbf{\hat{r}}_{\by}$ for any $ k \geq 2 $, it follows that $ \bZ_{j.j_1\dots j_{k}}^T \by $ can be written as a function of  $ \mathbf{\hat{\Sigma}} $ and $\mathbf{\hat{r}}_{\by}$ for any $ k \geq 2 $.
	This completes the proof of Lemma \ref{lemma:step1}. \hfill $ \square $
	
	\paragraph{Proof of Lemma 2.}
	The $ F $-statistic at step $ k \geq 2 $ of the forward stepwise search is given by 
	\begin{align*}
		F = \frac{\text{RSS}_{k-1} - \text{RSS}_{k}}{\text{RSS}_{k}} \times (n - k - 1),
	\end{align*}
	where $ F $ has $ \text{df}_1=1 $ and $ \text{df}_2=k $ degrees of freedom, respectively. Note that 
	\begin{align*}
		\text{RSS}_1 &= \left\lVert \by - \hat{r}_{y j_1} \bX_{j_1} \right\rVert_2^2 = 1 - \hat{r}_{yj_1}, \quad \text{and}\\
		\text{RSS}_k &= \left\lVert\by - \hat{r}_{y j_1} \bX_{j_1} - \sum_{r = 2}^{k} \beta_{y j_r.j_1 \dots j_{(r-1)}} \bZ_{j_r.j_1 \dots j_{(r-1)}}\right\rVert_2^2, \quad k \geq 2,
	\end{align*}
	where 
	\begin{align*}
		\beta_{y j_r.j_1 \dots j_{(r-1)}} = \frac{\bZ_{j_r.j_1 \dots j_{(r-1)}}^T \by}{\bZ_{j_r.j_1 \dots j_{(r-1)}}^T\bZ_{j_r.j_1 \dots j_{(r-1)}}}.
	\end{align*}
	Note that we may write
	\begin{align*}
		\text{RSS}_k &= \left\lVert\by - \hat{r}_{y j_1} \bX_{j_1} - \sum_{r = 2}^{k-1} \beta_{y j_r.j_1 \dots j_{(r-1)}} \bZ_{j_r.j_1 \dots j_{(r-1)}} - \beta_{y j_k.j_1 \dots j_{k-1}} \bZ_{j_k.j_1 \dots j_{k-1}}\right\rVert_2^2 \\
		&= \left\lVert \left(\by - \hat{r}_{y j_1} \bX_{j_1} - \sum_{r = 2}^{k-1} \beta_{y j_r.j_1 \dots j_{(r-1)}} \bZ_{j_r.j_1 \dots j_{(r-1)}} \right) - \beta_{y j_k.j_1 \dots j_{k-1}} \bZ_{j_k.j_1 \dots j_{k-1}}\right\rVert_2^2.
	\end{align*}
	By the orthogonality of $ \bX_{j_1} $ and the residual vectors $\bZ_{j_r.j_1 \dots j_{(r-1)}}$, $2 \leq r \leq k$, it follows that
	\begin{align*}
		\text{RSS}_k &= \text{RSS}_{k-1} - 2 \left[ \beta_{y j_k.j_1 \dots j_{k-1}} \by^T  \bZ_{j_k.j_1 \dots j_{k-1}}\right] + \beta_{y j_k.j_1 \dots j_{k-1}}^2 \bZ_{j_k.j_1 \dots j_{k-1}}^T  \bZ_{j_k.j_1 \dots j_{k-1}} \\
		&= \text{RSS}_{k-1} - 2 \left[ \frac{ \left(\by^T  \bZ_{j_k.j_1 \dots j_{k-1}}\right)^2}{\bZ_{j_k.j_1 \dots j_{k-1}}^T  \bZ_{j_k.j_1 \dots j_{k-1}}}\right] +  \frac{ \left(\by^T  \bZ_{j_k.j_1 \dots j_{k-1}}\right)^2}{\bZ_{j_k.j_1 \dots j_{k-1}}^T  \bZ_{j_k.j_1 \dots j_{k-1}}} \\
		&= \text{RSS}_{k-1} - \frac{\left(\bZ_{j_k.j_1 \dots j_{k-1}}^T \by\right)^2}{\bZ_{j_k.j_1 \dots j_{k-1}}^T\bZ_{j_k.j_1 \dots j_{k-1}}}.
	\end{align*}
	In the proof of Lemma \ref{lemma:step1} it was shown that the  numerator and denominator on the right-hand side can be written as a function of $ \mathbf{\hat{\Sigma}} $ and $\mathbf{\hat{r}}_{\by}$ only for any $ k \geq 2 $. This completes the proof of Lemma \ref{lemma:step2}. \hfill $ \square $
	
	\subsection*{Proof of Proposition 2}
	
	First note that for the first variable added in a model, it is trivial to show that that finding the optimal initial variable and computing its RSS only depends on $ \mathbf{\hat{r}_{y}} $ and $ \mathbf{\hat{\Sigma}} $. In particular, assuming the response and predictor variables have been scaled to have mean zero and variance one,
	\begin{align*}
		j_1 &= \argmax_{\substack{j \notin \mathcal{J}^g \\ 1 \leq g \leq G}} |\hat{r}_{yj}|, \\ \frac{1}{n}\text{RSS}_1 &= \frac{1}{n}\left(\by - \hat{r}_{yj_1} \bx_{j_1}\right)^T\left(\by - \hat{r}_{yj_1} \bx_{j_1}\right) \\
		&= \frac{1}{n}\left(\by^T \by - 2 \hat{r}_{yj_1} \bx_{j_1}^T \by + \hat{r}_{yj_1}^2 \bx_{j_1}^T \bx_{j_1}^T \right) \\
		&= \frac{1}{n}\left(n - 2n \hat{r}_{yj_1}^2 + n \hat{r}_{yj_1}^2  \right) \\
		&= \frac{1}{n}\left(n - 2n \hat{r}_{yj_1}^2 + n \hat{r}_{yj_1}^2  \right) \\
		&= 1 - \hat{r}_{yj_1}^2.
	\end{align*}
	
	Now for notational convenience, denote for $k \geq 2$:
	\begin{align*}
		P_{yj_k.j_1 \dots j_{k-1}} &= \bZ_{yj_k.j_1 \dots j_{k-1}}^T \by, \quad \text{and} \\
		P_{j_aj_b.j_1 \dots j_{k-1}} &= \bZ_{yj_k.j_1 \dots j_{k-1}}^T \bZ_{j_aj_b.j_1 \dots j_{k-1}}.
	\end{align*}
	
	At step $ k \geq 2$, variable the $k$-th optimal predictor $ j_k $ for a given model is given by  
	\begin{align*}
		j_k &= \argmax_{j \in \mathcal{J}\setminus\{j_1, \dots, j_{k-1}\}} \left| \frac{\bZ_{j.j_1\dots j_{k-1}}^T \by/n}{\sqrt{\bZ_{j.j_1\dots j_{k-1}}^T\bZ_{j.j_1\dots j_{k-1}}/n}} \right| \\
		&=  \argmax_{j \in \mathcal{J}\setminus\{j_1, \dots, j_{k-1}\}} \left| \frac{P_{yj_k.j_1 \dots j_{k-1}}}{\sqrt{P_{jj.j_1 \dots j_{k-1}}}} \right|.
	\end{align*}
	Also, from the proof of Lemma 2, it follows that 
	\begin{align*}
		\text{RSS}_k &= \text{RSS}_{k-1} - \frac{P_{yj.j_1 \dots j_{k-1}}^2}{P_{j_kj_k.j_1 \dots j_{k-1}}}.
	\end{align*}
	
	To complete the proof, we need to show that $P_{yj.j_1 \dots j_k}$ and $P_{j_aj_b.j_1 \dots j_k}$, for $k \geq 2 $, can be written as a function of $ \mathbf{\hat{r}_{y}} $ and $ \mathbf{\hat{\Sigma}} $ only. We begin with $P_{yj.j_1 \dots j_k}$. In the proof of Lemma 1, we already showed that
	\begin{align*}
		\frac{1}{n}\bZ_{j.j_1}^T \by = \hat{r}_{y j} - \mathbf{\hat{\Sigma}}_{j j_1} \hat{r}_{y j_1}, \quad j \in \mathcal{J}\setminus\{j_1, \dots, j_{k-1}\},
	\end{align*}
	thus it follows that 
	\begin{align*}
		P_{yj_k.j_1} = n \left( \hat{r}_{y j} - \mathbf{\hat{\Sigma}}_{j j_1} \hat{r}_{y j_1}\right), \quad j \in \mathcal{J}\setminus\{j_1, \dots, j_{k-1}\},
	\end{align*}
	Also, from the proof of Lemma 1,
	\begin{align*}
		P_{yj.j_1 \dots j_k} = P_{yj.j_1 \dots j_{k-1}} - \beta_{j j_k.j_1 \dots j_{k-1}} P_{yj_k.j_1 \dots j_{k-1}}, \quad j \in \mathcal{J}\setminus\{j_1, \dots, j_{k-1}\}.
	\end{align*}
	where
	\begin{align*}
		\beta_{j_aj_b.j_1\dots j_{k}}= \frac{\bZ_{j_a.j_1\dots j_{k}}^T\bZ_{j_b.j_1\dots j_{k}}}{\bZ_{j_{b}.j_1\dots j_{k}}^T\bZ_{j_{b}.j_1\dots j_{k}}}, \quad j_a, j_b \in \mathcal{J}\setminus\{j_1, \dots, j_{k}\}.
	\end{align*}
	putting it all together, it follows that 
	\begin{align*}
		P_{yj.j_1 \ldots j_k} = P_{yj.j_1 \ldots j_{k-1}} - \frac{P_{jj_k.j_1 \ldots j_{k-1}}}{P_{j_k j_k.j_1 \ldots j_{k-1}}} P_{yj_k.j_1 \ldots j_{k-1}}.
	\end{align*}
	Thus $P_{yj.j_1 \ldots j_k}$ can be written as a function of $ \mathbf{\hat{r}_{y}} $ and $ \mathbf{\hat{\Sigma}} $ alone. 
	
	Now, for $P_{j_aj_b.j_1 \dots j_k}$, recall from the proof of Lemma 1 that
	\begin{align*}
		\bZ_{j_a.j_1\dots j_{k}}^T \bZ_{j_b.j_1\dots j_{k}}  &= \bZ_{j_a.j_1\dots j_{k-1}}^T \bZ_{j_b.j_1\dots j_{k-1}} \\
		&\quad\quad - \beta_{j_b j_{k}.j_1\dots j_{k-1}} \bZ_{j_a.j_1\dots j_{k-1}}^T \bZ_{j_k.j_1\dots j_{k-1}} \\
		& \quad \quad - \beta_{j_a j_{k}.j_1\dots j_{k-1}} \bZ_{j_b.j_1\dots j_{k-1}}^T \bZ_{j_k.j_1\dots j_{k-1}} \\
		&\quad\quad + \beta_{j_a j_{k}.j_1\dots j_{k-1}} \beta_{j_b j_{k}.j_1\dots j_{k-1}}  \bZ_{j_{k}.j_1\dots j_{k-1}}^T \bZ_{j_{k}.j_1\dots j_{k-1}}.
	\end{align*}
	Thus putting it all together,
	\begin{align*}
		P_{j_a j_b.j_1 \ldots j_k} = P_{j_a j_b.j_1 \ldots j_{k-1}} - \frac{P_{j_a j_k.j_1 \ldots j_{k-1}} P_{j_b j_k.j_1 \ldots j_{k-1}}}{P_{j_k j_k.j_1 \ldots j_{k-1}}}.
	\end{align*}
	It follows that $P_{j_a j_b.j_1 \ldots j_k}$ for $k \geq 2$ can be written as a function of $ \mathbf{\hat{r}_{y}} $ and $ \mathbf{\hat{\Sigma}} $ alone. This completes the proof of Proposition 2. \hfill $ \square $
	
	\section*{Appendix E: Convergence Analysis of the PSBGD Algorithm} \label{sec:psbgd}
	
	In this section, we provide the theoretical justification for the Projected Subset Block Gradient Descent (PSBGD) algorithm presented in Algorithm 2 of the main article. We begin by formally defining the concepts required for the analysis, then derive the key properties of the objective function, and finally provide a detailed proof for Proposition 3.
	
	\subsection{Preliminaries and Lipschitz Constant Derivation}
	
	The core of our computing strategy involves solving the RMSS problem by iteratively updating blocks of variables. For each model $g, 1 \leq g \leq G$, the inner loop of Algorithm 2 performs a two-block coordinate descent on the coefficient vector $\bbet^g$ and the auxiliary residual vector $\boldsymbol{\eta}^g$. The objective function for a single model $g$, given the other models, is:
	\begin{equation}
		\mathcal{L}_n\left(\bbet^g, \boldsymbol{\eta}^g \vert \by, \bX\right) = \left\lVert \by - \bX \bbet^g  - \boldsymbol{\eta}^g  \right\rVert_2^2. \label{eq:single_model_loss}
	\end{equation}
	The gradients of $\mathcal{L}_n$ with respect to $\bbet^g$ and $\boldsymbol{\eta}^g$ are given by:
	\begin{align}
		\nabla_{\bbet} \mathcal{L}_n\left(\bbet^g, \boldsymbol{\eta}^g | \by, \bX \right) &= -2\bX^T \left(\by - \bX \bbet^g - \boldsymbol{\eta}^g \right), \label{eq:gradient_beta}\\ 
		\nabla_{\boldsymbol{\eta}} \mathcal{L}_n\left(\bbet^g, \boldsymbol{\eta}^g | \by, \bX \right) &= -2  \left(\by - \bX \bbet^g - \boldsymbol{\eta}^g\right). \label{eq:gradient_eta}
	\end{align}
	A function's gradient $\nabla f(\mathbf{z})$ is said to be Lipschitz continuous with constant $\ell$ if $\lVert \nabla f(\mathbf{z}_1) - \nabla f(\mathbf{z}_2) \rVert_2 \leq \ell \lVert \mathbf{z}_1 - \mathbf{z}_2 \rVert_2$ for all $\mathbf{z}_1, \mathbf{z}_2$. This property is crucial for guaranteeing the convergence of gradient-based methods.
	
	For the gradient with respect to $\bbet^g$, we consider two vectors $\bbet_1^g, \bbet_2^g$ that are both supported on the active set $\mathcal{S}_u^g$ (as enforced by the algorithm):
	\begin{align*}
		&\left\Vert \nabla_{\bbet}	\mathcal{L}_n\left(\bbet_1^g, \boldsymbol{\eta}^g | \by, \bX \right) - \nabla_{\bbet} \mathcal{L}_n\left(\bbet_2^g, \boldsymbol{\eta}^g | \by, \bX \right) \right\Vert_2 \\
		&= \left\Vert -2\bX^T \left(\by - \bX \bbet_1^g - \boldsymbol{\eta}^g \right) - (-2\bX^T \left(\by - \bX \bbet_2^g - \boldsymbol{\eta}^g \right)) \right\Vert_2 \\
		&= \left\Vert 2 \bX^T \bX \left(\bbet_1^g - \bbet_2^g \right) \right\Vert_2 
		= \left\Vert 2 \bX_{\mathcal{S}_u^g}^T \bX_{\mathcal{S}_u^g} \left((\bbet_1^g)_{\mathcal{S}_u^g} - (\bbet_2^g)_{\mathcal{S}_u^g} \right) \right\Vert_2 \\
		&\leq 2 \left\Vert \bX_{\mathcal{S}_u^g}^T \bX_{\mathcal{S}_u^g} \right\Vert_2 \left\Vert \bbet_1^g - \bbet_2^g \right\Vert_2.
	\end{align*}
	Thus, $\nabla_{\bbet} \mathcal{L}_n$ is block-wise Lipschitz continuous with constant $\ell_{\bbet^{(g)}} = 2 \lVert \bX_{\mathcal{S}_u^g}^T \bX_{\mathcal{S}_u^g} \rVert_2$, where $\lVert \cdot \rVert_2$ denotes the spectral norm. For the gradient with respect to $\boldsymbol{\eta}^g$:
	\begin{align*}
		& \left\lVert \nabla_{\boldsymbol{\eta}}	\mathcal{L}_n\left(\bbet^g, \boldsymbol{\eta}_1^g | \by, \bX \right) - \nabla_{\boldsymbol{\eta}} \mathcal{L}_n\left(\bbet^g, \boldsymbol{\eta}_2^g | \by, \bX \right)  \right\rVert_2\\
		&= \left\Vert -2\left(\by - \bX \bbet^g - \boldsymbol{\eta}_1^g \right) - (-2\left(\by - \bX \bbet^g - \boldsymbol{\eta}_2^g \right)) \right\Vert_2 \\
		&= \left\Vert 2 \left(\boldsymbol{\eta}_1^g - \boldsymbol{\eta}_2^g \right) \right\Vert_2 
		= 2 \left\Vert \boldsymbol{\eta}_1^g - \boldsymbol{\eta}_2^g \right\Vert_2.
	\end{align*}
	Thus, $\nabla_{\boldsymbol{\eta}} \mathcal{L}_n$ is Lipschitz continuous with constant $\ell_{\boldsymbol{\eta}} = 2$.
	
	\subsection*{Proof of Proposition 3}
	
	To prove the convergence and establish the rate for the PSBGD algorithm for a fixed model $g$, we analyze the inner loop (Step 1.2 of Algorithm 2 of the main article). This is a two-block coordinate descent procedure on the variables $\bbet^g$ and $\boldsymbol{\eta}^g$. For notational simplicity within this proof, we drop the superscript $g$ and denote the sequence of iterates for this model as $(\bbet_k, \boldsymbol{\eta}_k)$, with the objective function being $F(\bbet, \boldsymbol{\eta}) = \left\lVert \by - \bX \bbet - \boldsymbol{\eta} \right\rVert_2^2$. The updates at iteration $k$ are:
	\begin{align}
		\bbet_{k+1} &\in \mathcal{Q}\left(\bbet_k - \frac{1}{L_{\bbet}} \nabla_{\bbet} F(\bbet_k, \boldsymbol{\eta}_k); \, \mathcal{S}_u^{g}, t \right) \label{eq:proof_b_update} \\
		\boldsymbol{\eta}_{k+1} &\in \mathcal{P}\left(\boldsymbol{\eta}_k - \frac{1}{L_{\boldsymbol{\eta}}} \nabla_{\boldsymbol{\eta}} F(\bbet_{k+1}, \boldsymbol{\eta}_k); n - h \right) \label{eq:proof_eta_update}
	\end{align}
	The proof consists of three parts: establishing sufficient decrease for each block update, showing sequence convergence, and deriving the convergence rate.
	
	\paragraph{Part 1: Sufficient Decrease Property.}
	As derived in Section \ref{sec:psbgd}.A, the objective function $F$ is continuously differentiable and block-wise smooth. Its gradient $\nabla_{\bbet}F$ is Lipschitz continuous on the active set $\mathcal{S}_u^g$ with constant $\ell_{\bbet^{(g)}}$, and $\nabla_{\boldsymbol{\eta}}F$ is Lipschitz continuous with constant $\ell_{\boldsymbol{\eta}} = 2$. The algorithm uses step sizes $L_{\bbet} \geq \ell_{\bbet^{(g)}}$ and $L_{\boldsymbol{\eta}} \geq \ell_{\boldsymbol{\eta}}$.
	
	For the $\bbet$-update, by the standard descent lemma for Lipschitz continuous gradients, we have:
	\begin{equation}
		F(\bbet_{k+1}, \boldsymbol{\eta}_k) \leq F(\bbet_k, \boldsymbol{\eta}_k) + \nabla_{\bbet}F(\bbet_k, \boldsymbol{\eta}_k)^T(\bbet_{k+1} - \bbet_k) + \frac{L_{\bbet}}{2} \left\lVert \bbet_{k+1} - \bbet_k \right\rVert_2^2. \label{eq:proof_descent_lemma}
	\end{equation}
	The update \eqref{eq:proof_b_update} is a projection onto the non-convex set $\mathcal{C}_{\bbet} = \{\mathbf{w} : \lVert \mathbf{w} \rVert_0 \leq t, \text{supp}(\mathbf{w}) \subseteq \mathcal{S}_u^g\}$. A necessary first-order optimality condition for this projection is:
	$$
	\left( \bbet_{k+1} - \left(\bbet_k - \frac{1}{L_{\bbet}} \nabla_{\bbet}F(\bbet_k, \boldsymbol{\eta}_k)\right) \right)^T (\bbet_k - \bbet_{k+1}) \geq 0.
	$$
	Rearranging this inequality gives a bound on the gradient term:
	$$
	\nabla_{\bbet}F(\bbet_k, \boldsymbol{\eta}_k)^T(\bbet_{k+1} - \bbet_k) \leq -L_{\bbet} \left\lVert \bbet_{k+1} - \bbet_k \right\rVert_2^2.
	$$
	Substituting this back into \eqref{eq:proof_descent_lemma} yields a sufficient decrease:
	\begin{align}
		F(\bbet_{k+1}, \boldsymbol{\eta}_k) &\leq F(\bbet_k, \boldsymbol{\eta}_k) - L_{\bbet} \left\lVert \bbet_{k+1} - \bbet_k \right\rVert_2^2 + \frac{L_{\bbet}}{2} \left\lVert \bbet_{k+1} - \bbet_k \right\rVert_2^2 \nonumber \\
		&= F(\bbet_k, \boldsymbol{\eta}_k) - \frac{L_{\bbet}}{2} \left\lVert \bbet_{k+1} - \bbet_k \right\rVert_2^2. \label{eq:proof_decrease1}
	\end{align}
	A similar argument for the $\boldsymbol{\eta}$-update gives:
	\begin{equation}
		F(\bbet_{k+1}, \boldsymbol{\eta}_{k+1}) \leq F(\bbet_{k+1}, \boldsymbol{\eta}_k) - \frac{L_{\boldsymbol{\eta}}}{2} \left\lVert \boldsymbol{\eta}_{k+1} - \boldsymbol{\eta}_k \right\rVert_2^2. \label{eq:proof_decrease2}
	\end{equation}
	Combining \eqref{eq:proof_decrease1} and \eqref{eq:proof_decrease2}, we obtain the decrease over one full iteration:
	\begin{equation}
		F(\bbet_{k+1}, \boldsymbol{\eta}_{k+1}) \leq F(\bbet_k, \boldsymbol{\eta}_k) - \frac{L_{\bbet}}{2} \left\lVert \bbet_{k+1} - \bbet_k \right\rVert_2^2 - \frac{L_{\boldsymbol{\eta}}}{2} \left\lVert \boldsymbol{\eta}_{k+1} - \boldsymbol{\eta}_k \right\rVert_2^2. \label{eq:proof_total_decrease}
	\end{equation}
	
	\paragraph{Part 2: Convergence of the Sequence.}
	Inequality \eqref{eq:proof_total_decrease} shows that the sequence of objective function values $\{F(\bbet_k, \boldsymbol{\eta}_k)\}_{k=0}^{\infty}$ is non-increasing. Since $F(\bbet, \boldsymbol{\eta}) = \left\lVert \cdot \right\rVert_2^2 \geq 0$, the sequence is bounded below. A non-increasing sequence that is bounded below must converge.
	Summing \eqref{eq:proof_total_decrease} from $k=0$ to $K-1$ gives a telescoping series:
	$$
	\sum_{k=0}^{K-1} \left( \frac{L_{\bbet}}{2} \left\lVert \bbet_{k+1} - \bbet_k \right\rVert_2^2 + \frac{L_{\boldsymbol{\eta}}}{2} \left\lVert \boldsymbol{\eta}_{k+1} - \boldsymbol{\eta}_k \right\rVert_2^2 \right) \leq F(\bbet_0, \boldsymbol{\eta}_0) - F(\bbet_K, \boldsymbol{\eta}_K).
	$$
	As $K \to \infty$, the right-hand side is bounded by $F(\bbet_0, \boldsymbol{\eta}_0)$. This implies that the series of squared differences converges, which in turn requires the terms to approach zero:
	$$
	\lim_{k \to \infty} \left\lVert \bbet_{k+1} - \bbet_k \right\rVert_2 = 0 \quad \text{and} \quad \lim_{k \to \infty} \left\lVert \boldsymbol{\eta}_{k+1} - \boldsymbol{\eta}_k \right\rVert_2 = 0.
	$$
	This shows that the sequence of iterates stabilizes. Every limit point $(\bbet^*, \boldsymbol{\eta}^*)$ of the sequence $(\bbet_k, \boldsymbol{\eta}_k)$ is a stationary point of the algorithm, satisfying the fixed-point conditions stated in Proposition 4.
	
	\paragraph{Part 3: Convergence Rate.}
	To establish the $O(1/\delta)$ rate to a $\delta$-approximate stationary point, we rearrange \eqref{eq:proof_total_decrease}:
	$$
	\frac{L_{\bbet}}{2} \left\lVert \bbet_{k+1} - \bbet_k \right\rVert_2^2 + \frac{L_{\boldsymbol{\eta}}}{2} \left\lVert \boldsymbol{\eta}_{k+1} - \boldsymbol{\eta}_k \right\rVert_2^2 \leq F_k - F_{k+1},
	$$
	where $F_k = F(\bbet_k, \boldsymbol{\eta}_k)$. Summing from $k=0$ to $K-1$:
	$$
	\sum_{k=0}^{K-1} \left( \frac{L_{\bbet}}{2} \left\lVert \bbet_{k+1} - \bbet_k \right\rVert_2^2 + \frac{L_{\boldsymbol{\eta}}}{2} \left\lVert \boldsymbol{\eta}_{k+1} - \boldsymbol{\eta}_k \right\rVert_2^2 \right) \leq F_0 - F_K \leq F_0.
	$$
	The minimum value of the step-wise progress over the first $K$ iterations must be less than or equal to the average progress:
	$$
	\min_{0 \leq k < K} \left( \frac{L_{\bbet}}{2} \left\lVert \bbet_{k+1} - \bbet_k \right\rVert_2^2 + \frac{L_{\boldsymbol{\eta}}}{2} \left\lVert \boldsymbol{\eta}_{k+1} - \boldsymbol{\eta}_k \right\rVert_2^2 \right) \leq \frac{1}{K}\sum_{k=0}^{K-1} (\dots) \leq \frac{F_0}{K}.
	$$
	Let $C_{\min} = \min(L_{\bbet}/2, L_{\boldsymbol{\eta}}/2)$. Then we have:
	$$
	\min_{0 \leq k < K} \left( \left\lVert \bbet_{k+1} - \bbet_k \right\rVert_2^2 + \left\lVert \boldsymbol{\eta}_{k+1} - \boldsymbol{\eta}_k \right\rVert_2^2 \right) \leq \frac{F_0}{C_{\min} \cdot K}.
	$$
	The algorithm finds a $\delta$-approximate stationary point when $\left\lVert \bbet_{k+1} - \bbet_k \right\rVert_2^2 \leq \delta$ and $\left\lVert \boldsymbol{\eta}_{k+1} - \boldsymbol{\eta}_k \right\rVert_2^2 \leq \delta$. This is guaranteed for at least one $k \in \{0, \dots, K-1\}$ if the sum of these squared norms is less than or equal to $2\delta$. We can ensure this by finding a $K$ such that $\frac{F_0}{C_{\min} \cdot K} \leq 2\delta$. This condition is met if $K \geq \frac{F_0}{2C_{\min}\delta}$. Thus, the number of iterations required is $O(1/\delta)$. \hfill $\square$
	
	\section*{Appendix F: Optional Neighborhood Search Refinement} \label{sec:neighborhood}
	
	The primary computing algorithm described in the main article generates a grid of solutions by following a ``warm-start" path, primarily along the diversity parameter $u$. While computationally efficient, this greedy approach may settle in a local minimum. To further refine the solutions and explore the optimization landscape more thoroughly, we introduce an optional three-dimensional neighborhood search, detailed in Algorithm \ref{alg:neighborhood_search}.
	
	The core idea is to iteratively improve the solution for each grid point $(t_i, u_j, h_k)$ by re-running the optimization using its immediate neighbors in the grid as new warm starts. This allows information from better solutions to propagate across the grid. This process is repeated until a full pass over all grid points fails to produce a significant improvement in the total objective function value.

	\begin{algorithm}[ht!]
		\caption{Three-Dimensional Neighborhood Search \label{alg:neighborhood_search}}
		\begin{algorithmic}[1]
			\Require{Pairs $ (\hbbet^g[t, u, h], \hbet^g[t, u, h]) $, $ 1 \leq g \leq G $, for all combinations of $ t \in \mathcal{T} = \{t_1, \dots, t_q \} $, $ u \in \mathcal{U} = \{1, \dots, G\}$ and $ h \in \mathcal{H}  = \{h_1, \dots, h_r\}$, and tolerance parameter $\epsilon>0$.}
			\Statex
			\State For all possible $ (i, j, k) $ where $ i \in \{1, \dots, q\} $, $ j \in \{1, \dots, G\} $ and $ k \in \{1, \dots, r\} $: \label{alg4:step1}
			\begin{enumerate}
				\item[\footnotesize 1.1] Generate the neighborhood 
				\begin{align*}
					\mathcal{N}(i, j, k) = \{a \in \{1, \dots, q\}, b \in \{1, \dots, G\}, c \in \{1, \dots, r\}: |i - a| + |j - b| + |k - c| \leq 1\}.
				\end{align*}
				\item[\footnotesize 1.2] For all $ (a, b, c) \in \mathcal{N}(i, j, k) $:
				\begin{enumerate}
					\item[\footnotesize 1.2.1] Run Algorithm 2 of the main article initialized with $ (\hbbet^g[t_a, u_b, h_c], \hbet^g[t_a, u_b, h_c]) $, $ 1 \leq g \leq G $.
					\item[\footnotesize 1.2.2] Update $ (\hbbet^g[t_i, u_j, h_k], \hbet^g[t_i, u_j, h_k]) $, $ 1 \leq g \leq G $, with incumbent solution if it achieves a lower value of the objective function.
				\end{enumerate}
			\end{enumerate}
			\State Repeat step \ref{alg4:step1} until $ \epsilon $-small change is achieved for the sum 
			\begin{align*}
				\sum_{t \in \mathcal{T}} \sum_{u \in \mathcal{U}} \sum_{h \in \mathcal{H}} \sum_{g=1}^G \left\lVert \by - \bX \hbbet^g[t, u, h] - \hbet^g[t, u, h] \right\rVert_2^2
			\end{align*}
			\State Return the updated grid of solutions $ \hbbet^g[t, u, h] $, $ 1 \leq g \leq G $, for all combinations of $ t \in \mathcal{T} $, $ u \in \mathcal{U}$ and $ h \in \mathcal{H} $. 
		\end{algorithmic}
	\end{algorithm}

	While this neighborhood search can yield considerable improvements in minimizing the objective function, it significantly increases the overall computational cost of RMSS. Furthermore, based on our extensive numerical experiments, the improvements it provides in terms of out-of-sample prediction accuracy and variable selection stability are typically marginal compared to those from the more direct procedure described in the main article. For these reasons, we present it here as an optional refinement for users who prioritize finding the absolute best in-sample fit over computational speed.
	
	\section*{Appendix G: Simulation Tuning Parameters} \label{sec:tuning_specifications}
	
	This section details the specific tuning parameter settings used for the competing methods in our simulation study. All tuning parameters were selected using cross-validation as described in the main article.
	
	\begin{itemize}
		\item \textbf{Elastic Net (EN):} We used the \texttt{glmnet} package implementation. The $\ell_1$-$\ell_2$ mixing parameter was fixed at $\alpha=0.75$. The regularization parameter, $\lambda$, was selected from a log-spaced grid of 100 values automatically generated by the package's internal sequence.
		
		\item \textbf{PENSE and Adaptive PENSE:} We used the \texttt{pense} package. The EN mixing parameter was set to $\alpha=0.75$. To manage the method's computational expense, we reduced the number of initial candidates for its internal algorithm from the default. The penalty parameter $\lambda$ was chosen from a grid of 50 candidate values generated internally by the package. For Adaptive PENSE, a second cross-validation step was performed to determine the adaptive weights, as recommended by the authors.
		
		\item \textbf{Huber-EN:} We used the \texttt{hqreg} package. Consistent with the other penalized methods, the EN mixing parameter was set to $\alpha=0.75$. The penalty parameter $\lambda$ was selected from a grid of 50 values generated by the package.
		
		\item \textbf{SparseLTS:} We used the \texttt{robustHD} package. The penalty parameter $\lambda$ was selected from a log-equispaced grid of 50 values. This grid was generated using the \texttt{lambda0} function from the same package, which is designed to provide a reasonable search range for this method.
		
		\item \textbf{RMSS and RBSS:} For our proposed methods, the tuning parameters $(t, u, h)$ were selected via cross-validation over the specific grids described in the main article's simulation section.
	\end{itemize}
	
	\bibliographystyle{Chicago}
	\bibliography{Robust_Multi_Model_Subset_Selection}
	
\end{document}